\documentclass[apj]{emulateapj-rtx4}

\usepackage{color}
\usepackage{ulem}

\shorttitle{}
\shortauthors{Hashimoto et al.}

\begin{document}
\title{The Structure of Pre-transitional Protoplanetary Disks. II. \\
  Azimuthal Asymmetries, Different Radial Distributions of Large and Small Dust Grains in PDS~70.
  \footnote{Based on data collected at the Subaru Telescope, which is operated by the National Astronomical Observatory of Japan.}
  \footnote{The Submillimeter Array is a joint project between the Smithsonian Astrophysical Observatory and the Academia Sinica Institute of Astronomy and Astrophysics and is funded by the Smithsonian Institution and the Academia Sinica.}
}

  \author{
    J.~Hashimoto\altaffilmark{1}, 
    T.~Tsukagoshi\altaffilmark{2}, 
    J.~M.~Brown\altaffilmark{3}, 
    R.~Dong\altaffilmark{4,5}, 
    T.~Muto\altaffilmark{6}, 
    Z.~Zhu\altaffilmark{4,7}, 
    J.~Wisniewski\altaffilmark{1}, 
    N.~Ohashi\altaffilmark{8},
    T.~kudo\altaffilmark{8},
    N.~Kusakabe\altaffilmark{9},
    L.~Abe\altaffilmark{10}, 
    E.~Akiyama\altaffilmark{9},
    W.~Brandner\altaffilmark{11}, 
    T.~Brandt\altaffilmark{12}, 
    J.~Carson\altaffilmark{13,11},
    T.~Currie\altaffilmark{14}, 
    S.~Egner\altaffilmark{8}, 
    M.~Feldt\altaffilmark{11}, 
    C.~A.~Grady\altaffilmark{15,16}, 
    O.~Guyon\altaffilmark{8}, 
    Y.~Hayano\altaffilmark{8},
    M.~Hayashi\altaffilmark{9}, 
    S.~Hayashi\altaffilmark{8}, 
    T.~Henning\altaffilmark{11}, 
    K.~Hodapp\altaffilmark{17},
    M.~Ishii\altaffilmark{9},
    M.~Iye\altaffilmark{9}, 
    M.~Janson\altaffilmark{18,11},
    R.~Kandori\altaffilmark{9}, 
    G.~Knapp\altaffilmark{7}, 
    M.~Kuzuhara\altaffilmark{19},
    J.~Kwon\altaffilmark{20},
    T.~Matsuo\altaffilmark{21},
    M.~W.~McElwain\altaffilmark{16}, 
    S.~Mayama\altaffilmark{22}, 
    K.~Mede\altaffilmark{20},
    S.~Miyama\altaffilmark{23}, 
    J.-I.~Morino\altaffilmark{9},
    A.~Moro-Martin\altaffilmark{24,7}, 
    T.~Nishimura\altaffilmark{8}, 
    T.-S.~Pyo\altaffilmark{8},
    G.~Serabyn\altaffilmark{25}, 
    T.~Suenaga\altaffilmark{26},
    H.~Suto\altaffilmark{9}, 
    R.~Suzuki\altaffilmark{9},
    Y.~Takahashi\altaffilmark{26},
    M.~Takami\altaffilmark{27},
    N.~Takato\altaffilmark{8}, 
    H.~Terada\altaffilmark{8}, 
    C.~Thalmann\altaffilmark{28,29}, 
    D.~Tomono\altaffilmark{8},
    E.~L.~Turner\altaffilmark{7,30}, 
    M.~Watanabe\altaffilmark{31}, 
    T.~Yamada\altaffilmark{32}, 
    H.~Takami\altaffilmark{9},
    T.~Usuda\altaffilmark{9},
    M.~Tamura\altaffilmark{20,9}
    }
  
  \altaffiltext{1}{Department of Physics and Astronomy, The University of Oklahoma, 440 W. Brooks St. Norman, OK 73019 USA; jun.hashimoto@ou.edu}
  \altaffiltext{2}{College of Science, Ibaraki University, Bunkyo 2-1-1, Mito 310-8512, Japan}
  \altaffiltext{3}{Harvard-Smithsonian Center for Astrophysics, 60 Garden St., MS 78, Cambridge, MA 02138, USA}
  \altaffiltext{4}{Hubble Fellow}
  \altaffiltext{5}{Astronomy Department, University of California, Berkeley, CA 94720, USA}
  \altaffiltext{6}{Division of Liberal Arts, Kogakuin University, 1-24-2, Nishi-Shinjuku, Shinjuku-ku, Tokyo, 163-8677, Japan}
  \altaffiltext{7}{Department of Astrophysical Sciences, Princeton University, Princeton, NJ, 08544, USA}
  \altaffiltext{8}{Subaru Telescope, 650 North A'ohoku Place, Hilo, HI 96720, USA}
  \altaffiltext{9}{National Astronomical Observatory of Japan, 2-21-1 Osawa, Mitaka, Tokyo 181-8588, Japan}
  \altaffiltext{10}{Laboratoire Hippolyte Fizeau, UMR6525, Universite de Nice Sophia-Antipolis, 28, avenue Valrose, 06108 Nice Cedex 02, France}
  \altaffiltext{11}{Max Planck Institute for Astronomy, K\"{o}nigstuhl 17, D-69117 Heidelberg, Germany}
  \altaffiltext{12}{Astrophysics Department, Institute for Advanced Study, Princeton, USA}
  \altaffiltext{13}{Department of Physics and Astronomy, College of Charleston, 58 Coming St., Charleston, SC29424, USA}
  \altaffiltext{14}{Department of Astronomy and Astrophysics, University of Toronto, 50 St. George Street, Toronto, ON, Canada}
  \altaffiltext{15}{Eureka Scientific, 2452 Delmer, Suite 100, Oakland CA 96002, USA}
  \altaffiltext{16}{Exoplanets and Stellar Astrophysics Laboratory, Code 667, Goddard Space Flight Center, Greenbelt, MD 20771 USA}
  \altaffiltext{17}{University of Hawaii, 640 North A'ohoku Place, Hilo, HI 96720, USA}
  \altaffiltext{18}{Astrophysics Research Center, Queen's University Belfast, Belfast, UK}
  \altaffiltext{19}{Department of Earth and Planetary Sciences, Tokyo Institute of Technology, 2-12-1 Ookayama, Meguro-ku, Tokyo 152-8551, Japan}
  \altaffiltext{20}{Department of Astronomy, The University of Tokyo, Hongo 7-3-1, Bunkyo-ku, Tokyo 113-0033, Japan}
  \altaffiltext{21}{Department of Astronomy, Kyoto University, Kita-shirakawa-Oiwake-cho, Sakyo-ku, Kyoto 606-8502, Japan}
  \altaffiltext{22}{The Graduate University for Advanced Studies (SOKENDAI), Shonan International Village, Hayama-cho, Miura-gun, Kanagawa 240-0193, Japan}
  \altaffiltext{23}{Hiroshima University, 1-3-2, Kagamiyama, Higashi-Hiroshima 739-8511, Japan}
  \altaffiltext{24}{Department of Astrophysics, CAB - CSIC/INTA, 28850 Torrej'on de Ardoz, Madrid, Spain}
  \altaffiltext{25}{Jet Propulsion Laboratory, California Institute of Technology, 4800 Oak Grove Drive, Pasadena, CA 91109, USA}
  \altaffiltext{26}{Department of Astronomical Science, Graduate University for Advanced Studies (Sokendai), Tokyo 181-8588, Japan}
  \altaffiltext{27}{Institute of Astronomy and Astrophysics, Academia Sinica, P.O. Box 23-141, Taipei 10617, Taiwan}
  \altaffiltext{28}{Institute for Astronomy, ETH Zurich Wolfgang-Pauli-Strasse 27, 8093 Zurich, Switzerland}
  \altaffiltext{29}{Astronomical Institute `Anton Pannekoek', University of Amsterdam, Science Park 904, 1098 XH Amsterdam, The Netherlands}
  \altaffiltext{30}{Kavli Institute for the Physics and Mathematics of the Universe, The University of Tokyo, Kashiwa 227-8568, Japan}
  \altaffiltext{31}{Department of Cosmosciences, Hokkaido University, Sapporo 060-0810, Japan}
  \altaffiltext{32}{Astronomical Institute, Tohoku University, Aoba, Sendai 980-8578, Japan}

  \begin{abstract}    
    The formation scenario of a gapped disk, i.e., transitional disk, 
    and its asymmetry 
    is still under debate. Proposed scenarios such as disk-planet interaction, photoevaporation, grain growth, 
    anticyclonic vortex, eccentricity,
    and their combinations
    would result in different radial distributions of the gas and the small (sub-$\mu$m size) and large (millimeter size) dust grains
    as well as asymmetric structures in a disk. 
    Optical/near-infrared (NIR) imaging observations and (sub-)millimeter interferometry can trace small and large dust grains,
    respectively;
    therefore multi-wavelength observations could help elucidate the origin of complicated structures of a disk. 
    Here we report SMA observations of the dust continuum at 1.3~mm and $^{12}$CO~$J=2\rightarrow1$ line emission
    of the pre-transitional protoplanetary disk around the solar-mass star PDS~70. PDS~70, a weak-lined T Tauri star,
      exhibits a gap in the scattered light from its disk
    with a radius of $\sim$65~AU at NIR wavelengths. However, we found a larger gap in the disk with a radius of $\sim$80~AU 
    at 1.3~mm. 
      Emission from
    all three disk components (the gas and the small and large dust grains) in images exhibits a deficit 
    in brightness in the central region of the disk, 
    in particular, the dust-disk in small and large dust grains has asymmetric brightness. 
    The contrast ratio of the flux density in the dust continuum between the peak position to the opposite side of the disk
    reaches 1.4. 
    We suggest 
    the asymmetries and different gap-radii of the disk around PDS~70 are potentially formed 
    by several (unseen) accreting planets inducing dust filtration.
  \end{abstract}
  \keywords{planetary systems --- protoplanetary disks --- stars: individual (PDS~70) 
    --- stars: pre-main sequence --- submillimeter: general --- polarization}

  \section{Introduction}\label{sec:intro}
  It is widely believed that protoplanetary disks dissipate inside-out \citep[c.f.,][]{will11} 
  and evolve into planetary systems \citep[e.g.,][]{haya85}. When an inner disk at less than $r\sim$1 AU begins to dissipate, 
  thermal emission from the disk significantly decreases at near- (NIR) to mid-infrared (MIR) wavelengths corresponding to 
  a blackbody temperature of $\sim$100-1000~K.  As a result, a deficit of flux density at $\sim$10 $\mu$m appears in an object's spectral 
  energy distribution (SED).  Such objects were first identified with the $IRAS$ satellite \citep{stro89} and are termed  
  `transitional disks (in transition between gas-rich primordial disks and gas-poor debris disks)' \citep{calv05,espa14}.
  Recent high resolution and high dynamic range imaging has directly revealed gaps in the central region of disks
  among roughly two dozen transitional disks at NIR \citep[e.g.,][]{thal10,hash12,maya12,debe13,garu13,quan13} 
  and (sub)-millimeter wavelengths \citep[e.g.,][]{piet05,brow09,andr11,will11,fuka13,pere14}. 
  Some transitional disks also show asymmetries in their dust continuum \citep[e.g.,][]{brow09,pere14}.
  To explain how the gap and the asymmetry form,
  a number of scenarios have been proposed: disk-planet interaction \citep{papa07,zhu11,dods11}, 
  photoevaporation\citep{clar01,alex06,alex07}, grain growth \citep{dull05,birn12},
  anticyclonic vortex \citep{rega12}, eccentricity \citep{kley06,atai13},
  and their combinations 
  \citep[dust filtration; radiation pressure;][]{rice06,dong12a,zhu12,pini12,owen14}. 
  
  Different gap formation scenarios could result in different radial distributions of the surface density 
  of the gas and dust grains. 
  For instance, {\it disk-planet interactions} will open a gap in the disk, but will have little effect 
  on the inner gas-disk \citep[e.g.,][]{zhu11}. Optically thick inner and outer disks in the optical/NIR wavelengths 
  would be separated by an optically thin gap.
  {\it Photoevaporation} should remove almost all the gas at the inner disk quickly, leaving a gas free cavity \citep[e.g.,][]{alex07}. 
  A supply of fresh material from the outer disk through mass accretion could be halted by a photoevaporative wind,
  resulting in remaining dust grains in the inner disk rapidly draining onto the star and creating a cavity
  (containing empty or optically thin materials). 
  {\it Grain growth} would mainly affect the dust grains, leaving gas unaffected \citep[e.g.,][]{dull05}.
  \citet{birn12} demonstrate that grain growth could account for the deficit of NIR and MIR excess in the SED of 
  transitional disks, whereas they predict large (millimeter size) dust grains still remain
  in the inner disk region. 
  Another intriguing scenario generating different radial distributions
  could be {\it dust filtration} \citep[e.g.,][]{rice06}, whereby large dust grains are trapped at the gas-pressure-bump,
  while small dust grains mixed well
  with the gas still inflow onto the central star.
  {\it Radiation pressure} from an accreting planet \citep{owen14} can also influence its nearby environment. When a massive planet 
  ($\gtrsim$3-4~$M_{\rm Jup}$) opens a gap and has sufficient mass accretion from the outer disk, radiation pressure
  from the planet is expected to hold back small dust grains, allowing the dust-free gas to accrete onto the central star.
  On the other hand, {\it anticyclonic vortices} \citep{rega12} and {\it eccentricities} \citep{kley06}
  result in asymmetric structures in the disk.  Anticyclonic vortices induced at the steep gas-pressure gradient such as a `dead zone' \citep{gamm96} and a gap-edge sculpted by
  planet(s) have a higher gas surface-density and efficiently gather dust particles into central vortices; 
  eccentric disks formed by a massive planet create a `traffic jam' of the gas and dust grains at the disk apocenter and 
  result in steady asymmetric features.

  Determining the radial and azimuthal 
  distributions of these disk components (the gas and the small and large dust grains) are crucial
  to understanding the formation of gaps in these disks as well as the origin of observed asymmetries. 
  However, transitional disks with a spatially resolved gap in all three disk components
  are relatively rare: AB~Aur \citep{piet05,hash11,tang12}; RX~J1604.3$-$2130~A \citep{math12,maya12}; SAO~206462 
  \citep[e.g.,][]{brow09,lyo11,garu13,pere14}; HD~142527 \citep[e.g.,][]{fuka06,casa13}; 
  Oph~IRS~48 \citep[e.g.,][Follette et~al. in prep.]{geer07,brow12,vand13,brud14}.

  Here we report the results of the dust continuum at 1.3~mm and $^{12}$CO~(2~$\rightarrow$~1) line emission
  from the pre-transitional protoplanetary disk around PDS~70 obtained with the Submillimeter Array (SMA).
  PDS~70 is a solar-type star with a mass of $\sim$0.8 $M_{\odot}$ at an assumed distance of 140~pc \citep{riau06}.
  The other basic properties of PDS~70 can be found in \citet{hash12}, \citet{dong12b}, and references therein. 
    The PDS~70 disk exhibits a large gap
  with a radius of $\sim$65 AU at NIR wavelengths \citep{hash12,dong12b}, which
  enables SMA observations to resolve a large gap in both dust continuum and the CO gas.
    In this paper, we discuss the gap formation scenario and the possible origin of fine structures in the disk 
    around PDS~70.

  \section{Observations and Results}\label{sec:obs}

  PDS~70 \citep[14:08:10.125, $-$41:23:52.81;][]{cutr13} was observed with the Submillimeter
  Array in three configurations in late 2012 and early 2013 (see Table~\ref{table:pds70obs}). 
  The receivers were tuned to place $^{12}$CO~(2~$\rightarrow$~1) in the
  upper sideband, although the exact tuning varied slightly between the
  tracks. The $^{12}$CO~(2~$\rightarrow$~1) line was observed with a resolution of 0.2~MHz in
  the compact and extended configuration, and a resolution of
  0.4~MHz in the very extended configuration. All tracks were gain
  calibrated with 1427-421 and 1325-430 and bandpass calibrated with
  3C~279. 
  The absolute flux scale was set by Titan and has a $\sim$10\%
  uncertainty. The data were calibrated in MIR following standard
  procedures\footnote{http://cfa-www.harvard.edu/\~{}cqi/mircook.html} and
  then combined and imaged in Miriad \citep{saul95}. The continuum data
  were concatenated from all chunks with 3.25 MHz resolution, providing
  $\sim$7.5 GHz of continuum bandwidth. The rms in the synthesized beam in
  the very extended configuration are 0.35~mJy~beam$^{-1}$ and 162.5~mJy~km~s$^{-1}$~beam$^{-1}$
  in the dust continuum and $^{12}$CO~(2~$\rightarrow$~1) line emissions, respectively.

\begin{figure}
   \centering
   \includegraphics[angle=0,scale=.85]{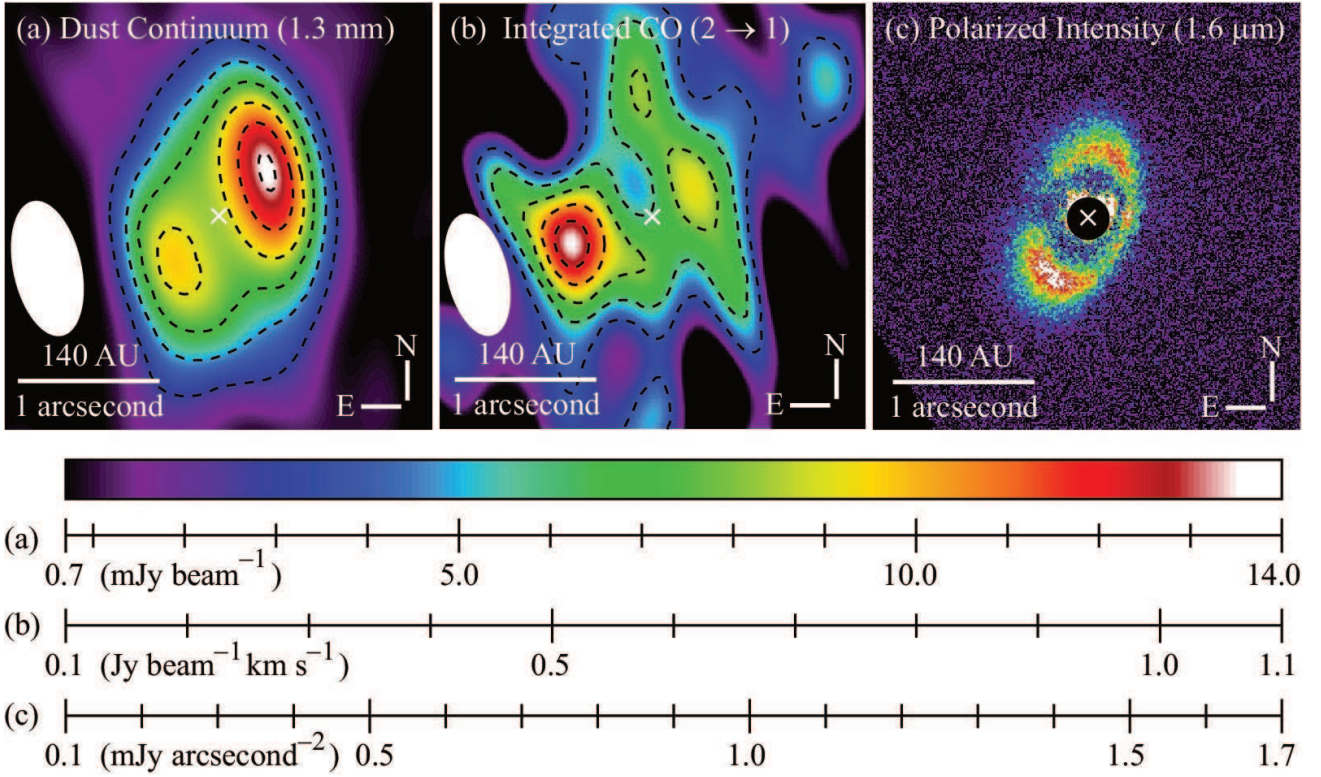}
   \caption{Observational results of PDS~70.
     (a): The dust continuum image at 1.3~mm.
     The beam size is $0\farcs97\times0\farcs48$ at PA~$=$~12.33$^{\circ}$ in the natural weight.
     Contours indicate 8, 14, 20, 26, 32, and 38~$\sigma$ (1~$\sigma =$0.35~mJy~beam$^{-1}$).
     (b): The integrated $^{12}$CO~(2~$\rightarrow$~1) emission image.
     The beam size is $0\farcs88\times0\farcs43$ at PA~$=$~13.41$^{\circ}$ in the natural weight.
     Contours indicate 2, 3, 4, 5, and 6~$\sigma$ (1~$\sigma =$162.5~mJy~km~s$^{-1}$~beam$^{-1}$).
     (c): The 1.6~$\mu$m polarized intensity image at the spatial resolution of $\sim$$0\farcs1$
     \citep{hash12}. Central region with a radius of $0\farcs1$
     is masked due to the dominant stellar photon noise. Crosses in each panels represent the astrometric point
     \citep[14:08:10.125, $-$41:23:52.81;][]{cutr13}.
   }\label{res:dust_pds70}
 \end{figure}

  The dust continuum image at 1.3~mm and the integrated $^{12}$CO~(2~$\rightarrow$~1) line emission map (natural weighting) 
  in the very extended configuration as well as the 1.6~$\mu$m polarized intensity image of PDS~70 are shown in 
  Fig.~\ref{res:dust_pds70}. The images of the dust continuum and $^{12}$CO~(2~$\rightarrow$~1) emissions in the very extended 
  configuration have a peak flux of 13.6~mJy~beam$^{-1}$ and 1.06~Jy~km~s$^{-1}$~beam$^{-1}$, respectively. Meanwhile,
  in the combined images of the compact; extended; very extended configurations, a total flux of the dust continuum and integrated 
  line emissions above 3~$\sigma$ has 38.1~$\pm$~1.1~mJy and 3.5~$\pm$~0.4~Jy~km~s$^{-1}$, respectively. 
  All three images in Fig.~\ref{res:dust_pds70} show
  a deficit of flux density in the central regions of the disk, 
    suggesting a
  central cavity or gap structures in small and large dust grains and the gas. 
We note that the NIR image alone does not provide convincing proof for the deficit of dust grains in the central region of the disk \citep{taka14}. \citet{taka14} shows that the such a deficit in the NIR flux still allows the possibility of a disk surface that is parallel to the light path from the star, or a disk that is shadowed by structures in the inner radii.

  The morphological structures in the dust continuum and NIR polarized intensity are asymmetric:
    $1.38 \pm 0.06$ and $1.32 \pm 0.15$ times brighter in the north-west part of the dust continuum and the south-east part of 
    NIR polarized intensity, respectively, while the brightness of the CO~(2~$\rightarrow$~1) map exhibits no clear asymmetries
      within the limits of observational errors (i.e., it is
    $1.43 \pm 0.38$ times brighter in the south-east region).
  Note that although the orientation of the disk-major axis in the dust continuum image (Fig.~\ref{res:dust_pds70}a) and 
  the NIR polarized intensity image (Fig.~\ref{res:dust_pds70}c) appear different from each other (PA of 135$^{\circ}$ and 
  160$^{\circ}$ respectively), this is due to the elongated beam in the SMA observations.

\begin{figure}
   \centering
   \includegraphics[angle=0,scale=1.0]{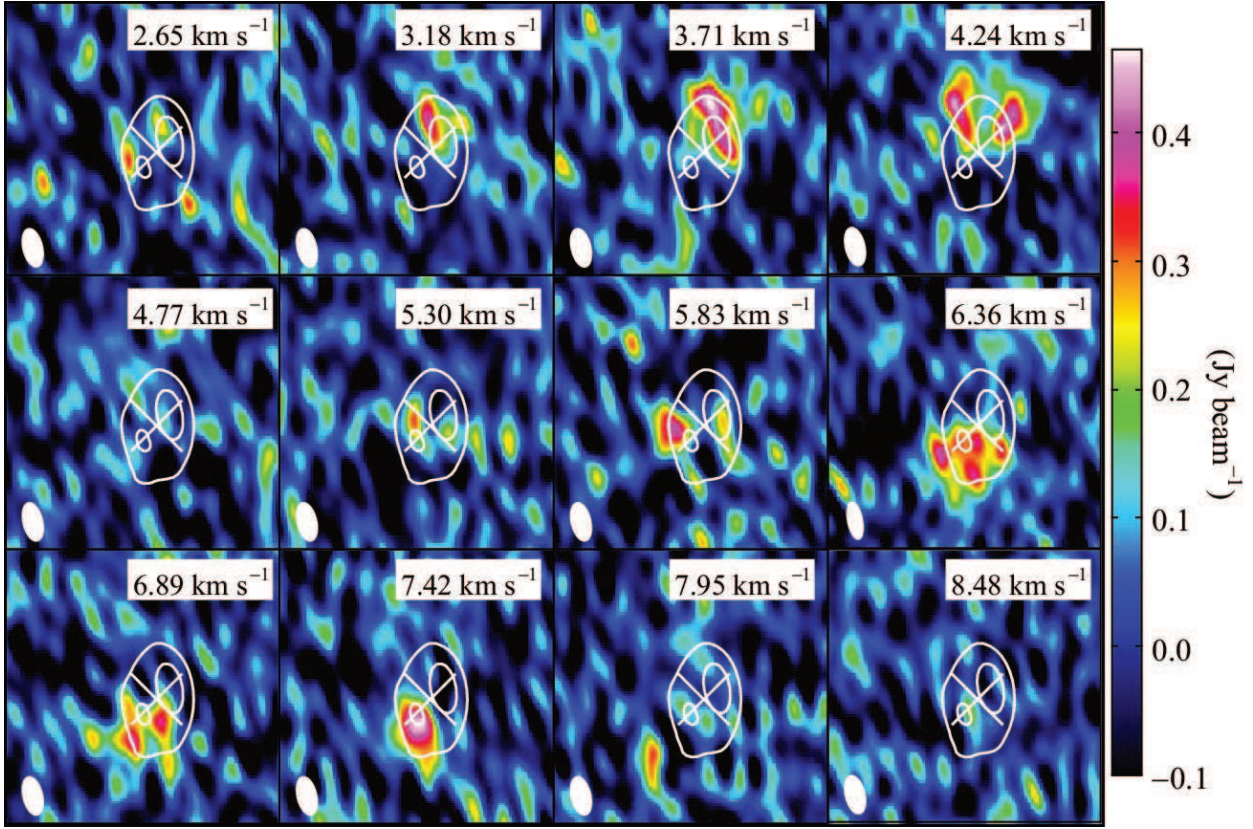}
   \caption{
     Channel maps of $^{12}$CO~(2~$\rightarrow$~1) line emission with natural weighting in the velocity range between 
     2.65 and 8.48~km~s$^{-1}$. The contours indicate the dust continuum in Fig.~\ref{res:dust_pds70}(a) with 8 and 26~$\sigma$ (
     1~$\sigma =$0.35~mJy~beam$^{-1}$). The synthesized beam is shown at the lower-left corner in each panel with
     $0\farcs88\times0\farcs43$ at PA~$=$~13.41$^{\circ}$. Crosses in each panels represent the astrometric point
     \citep[14:08:10.125, $-$41:23:52.81;][]{cutr13}.
   }\label{res:vel}
 \end{figure}

\begin{figure}
   \centering
   \includegraphics[angle=0,scale=.85]{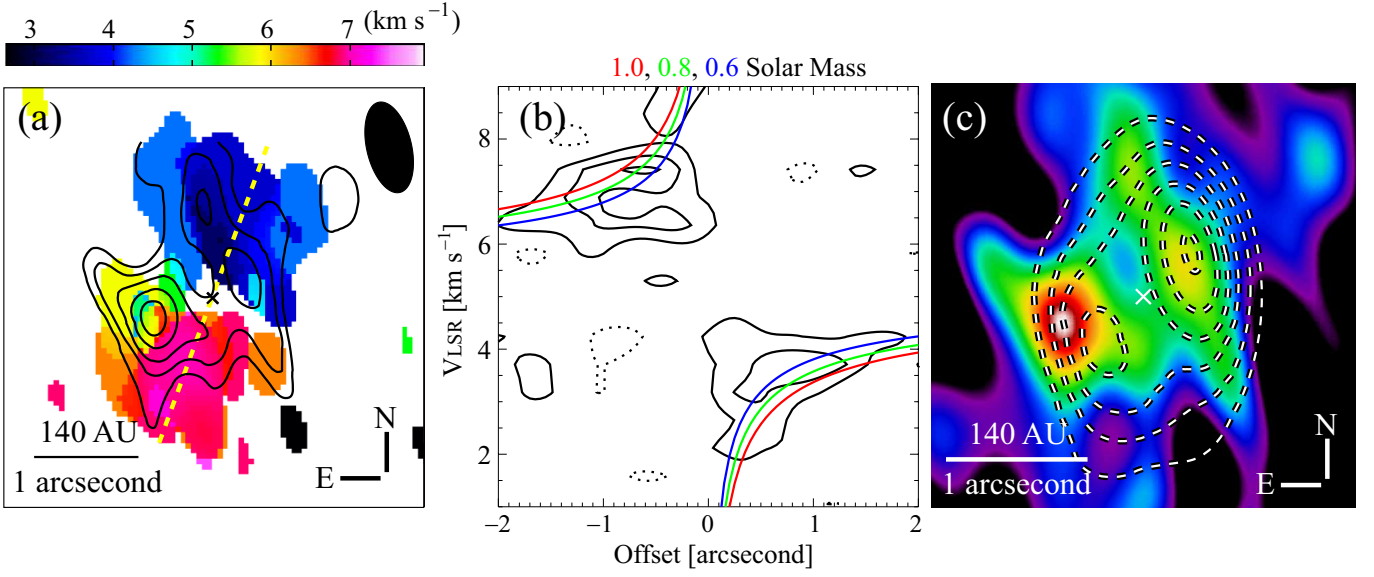}
   \caption{
     Observational results of $^{12}$CO~(2~$\rightarrow$~1) line emission of PDS~70.
     (a): Moment map with contours of the integrated $^{12}$CO~(2~$\rightarrow$~1) emission (taken from Fig.~\ref{res:dust_pds70}b;
     the beam size of $0\farcs88\times0\farcs43$ at PA~$=$~13.41$^{\circ}$). 
     (b): Position-velocity diagram along the yellow dashed-line (PA~$=$~160$^{\circ}$; centered at the astrometric point) 
     in panel (a). The disk inclination is set to 50$^{\circ}$ taken from \citet{hash12}. Colored thick-lines are loci of 
     peak emission in the Keplerian disk around the central star with a mass of 1.0 (red), 0.8 (green), and 0.6~$M_{\odot}$ (blue).
     Black contours are spaced by 2~$\sigma =$0.07~Jy~beam$^{-1}$.
     (c): The integrated $^{12}$CO~(2~$\rightarrow$~1) emission-line image with contours of the dust continuum taken from
     Fig.~\ref{res:dust_pds70}(a) and (b). Crosses in panels (a) and (c) represent the astrometric point
     \citep[14:08:10.125, $-$41:23:52.81;][]{cutr13}.
   }\label{res:gas}
 \end{figure}

  Fig.~\ref{res:vel} shows channel maps of the $^{12}$CO~(2~$\rightarrow$~1) line emission with the LSR velocity range from 
  2.65~km~s$^{-1}$ to 8.48~km~s$^{-1}$. 
    Since the $^{12}$CO line could easily turn out to be optically thick, we can estimate the gas temperature 
    based on the line intensity. The peak flux in the channel maps is 0.46~Jy~beam$^{-1}$ at 7.42~km~s$^{-1}$ and 
    suggests $\sim$28~K with the Rayleigh-Jeans approximation.
  A velocity gradient from the northwest to southeast can be seen in the channel maps (Fig.~\ref{res:vel}),
  suggestive of rotation. This velocity gradient can also be seen in the intensity-weighted mean velocity map
  (Fig.~\ref{res:gas}a). Assuming PDS~70 is surrounded by the Keplerian disk with an inclination of 50$^{\circ}$ \citep{hash12},
  we infer a dynamical mass of the central star with 0.6-0.8~$M_{\odot}$ in the position-velocity diagram (Fig.~\ref{res:gas}b), 
  consistent with the value of 0.82~$M_{\odot}$ estimated by the spectral type of K5 \citep{riau06}. 
  We also found the peak positions between the dust continuum image and the line-emission map are apparently 
  slightly misaligned in Fig.~\ref{res:gas}(c): the northwest parts are roughly aligned, 
  while the southeast parts have a misalignment with $\sim$$0\farcs25$ (35~AU).
  However, the peak-position determination accuracy strongly depends on the peak flux-density and the signal-to-noise ratio;
  hence, this misalignment seen in our noisy CO data may not be real.

  \section{Radiative Transfer Modeling}\label{sec:model} 

  In our previous studies \citep{hash12,dong12b}, we modeled the SED and NIR polarized intensity of PDS~70
  using a Monte Carlo radiative transfer (MCRT) code \citep{whit13} to derive the morphological structure of the dust disk,
  especially in small dust grains. Whitney's code follows a two-layer disk model with small (up to $\sim$$\mu$m size) 
  dust grains in the upper disk-atmosphere and large (up to millimeter size) dust grains in the disk mid-plane \citep[e.g.,][]{dale06}. 
  Historically, numerical simulations of a disk containing only small dust grains failed to reproduce millimeter-wave emission 
  in typical T~Tauri~stars due to the low opacity of small dust grains at longer wavelengths \citep{dale99}, 
  whereas a disk with only large dust grains was predicted to have no silicate emission bands \citep{dale01} 
  ubiquitously seen in T~Tauri~stars \citep[e.g.,][]{furl09} due to the gray opacity of large dust grains at shorter wavelengths.
  Thus, a two-layer disk model has been employed to improve the agreement with observations 
  \citep[e.g.,][]{chia01}. 

  In this paper, we attempt to simultaneously derive the disk properties in small and large dust grains 
  by fitting the SED, the polarized intensity image at NIR, and the dust-continuum image at 1.3~mm using Whitney's code.
  Since we mainly focus on the radial distributions of small and large dust grains (especially the radius of the gap)
  in the modeling efforts, 
  we attempt to derive the gap-radius by comparing radial profiles of synthesized images and observations.
  Note that the gas is not modeled in this paper.

  \subsection{Dust Disk Structure and Dust Property}\label{sec:struct}

\begin{figure}
   \centering
   \includegraphics[angle=0,scale=1.0]{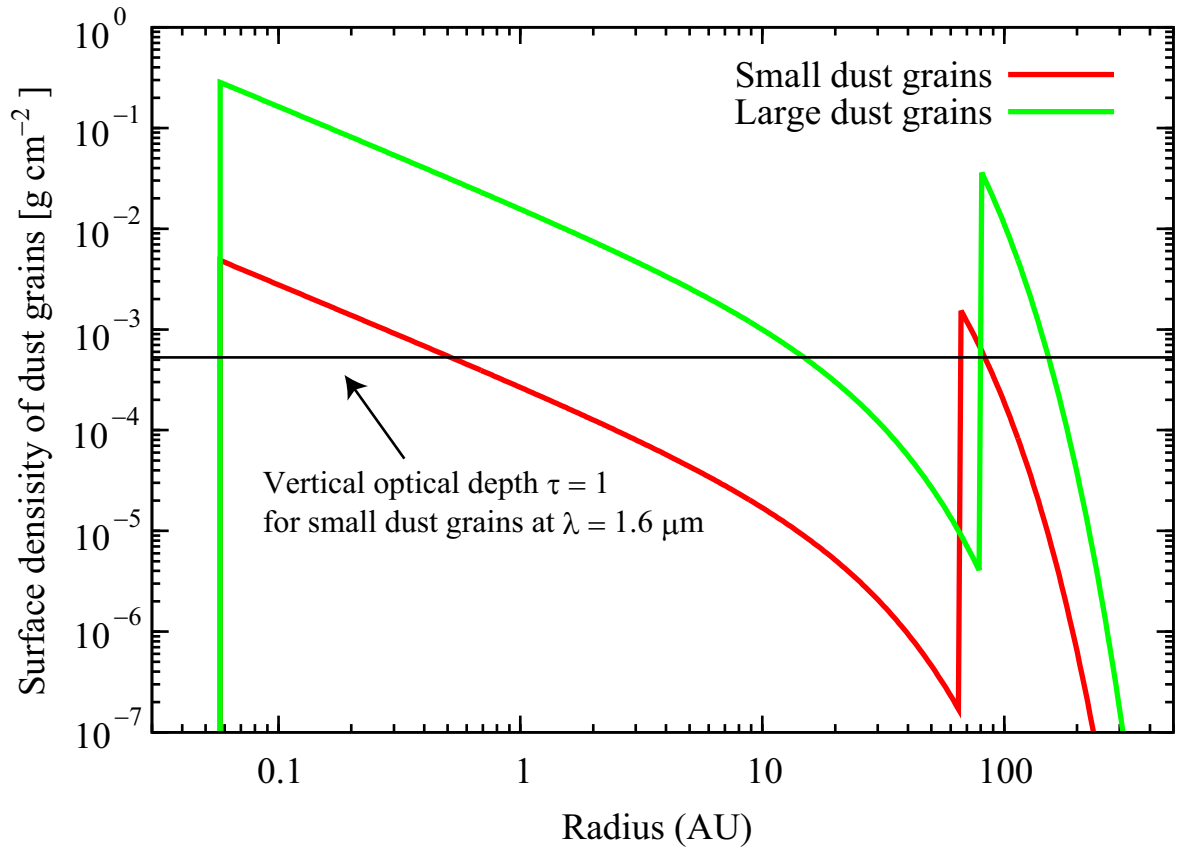}
   \caption{
     A surface density of small and large dust grains in our fiducial model.
   }\label{density}
 \end{figure}

  The dust-disk structure assumed in this paper follows that from  \citet{hash12} and \citet{dong12b}.
  We synthesize the SED, the NIR polarized intensity image, and the dust continuum at 1.3~mm utilizing 
  an axisymmetric and flared disk model in which the vertical density structure ($\rho$ in cylindrical-polar 
  coordinates \{$R$, $z$\}) in both small and large dust grains is assumed to be Gaussian,
  \begin{eqnarray}
     \rho(R,z) = \frac{\Sigma(R)}{\sqrt{2\pi}h} \ {\rm exp}\left[-\frac{1}{2}\left(\frac{z}{h}\right)^{2}\right],
     \label{eq_dens}
  \end{eqnarray}
  where $\Sigma$ is a surface density; $h$ is a scale height varying as a power law with a radius, i.e., $h \propto R^{p}$. 
  To reduce the number of parameters and simplify, we assume $p=1.25$\footnote{A scale height is described as 
    $h \approx c_{\rm s} \Omega^{-1}$, where $\Omega$ is the rotational angular velocity of $\Omega \propto R^{-1.5}$ in the Kepler 
    rotation; $c_{\rm s}$ is the sound speed of $c_{\rm s} \propto T^{0.5}$. If $T \propto R^{-0.5}$, then $h \propto R^{1.25}$.}, 
  i.e., a mid-plane temperature of $T \propto R^{-0.5}$ as approximately estimated by millimeter observations \citep{beck90}.
  The scale heights of small and large dust grains are generally different, 
  because small dust grains are expected to be coupled well
  with the gas and might have a similar scale height of the gas whereas large dust grains de-coupled with the gas
  could be likely to settle in the disk mid-plane due to the stellar gravity.
  A radial surface density structure ($\Sigma$) in dust grains described in the similarity solution for viscous accretion
  disks \citep{lynd74,hart98} is
  \begin{eqnarray}
     \Sigma(R) = \Sigma_{0} \left(\frac{R}{R_{\rm c}}\right)^{-q}{\rm exp}\left[-\left(\frac{R}{R_{\rm c}}\right)^{2-q}\right],
     \label{eq_sig}
  \end{eqnarray}
  where $\Sigma_{0}$ is a normalized surface density determined by a total disk mass ($M_{\rm disk}$); $R_{\rm c}$ is a characteristic radius;
  $q$ is a radial gradient parameter. In Whitney's code, in order to generate a gap structure in a central region of a disk,
  a surface density of a dust disk inside the gap ($R_{\rm gap}$) is uniformly scaled down as follows,
  \begin{eqnarray}
     \Sigma(R) = \delta \ \Sigma_{0} \left(\frac{R}{R_{\rm c}}\right)^{-q}{\rm exp}\left[-\left(\frac{R}{R_{\rm c}}\right)^{2-q}\right],
     \ {\rm at} \ R \lesssim R_{\rm gap},
     \label{eq_sig_d}
  \end{eqnarray}
  where $\delta$ is a constant depletion factor. Note that we assume $q=1$ as derived in the viscous disk model with 
  $T \propto R^{-0.5}$ \citep{hart98}. A radial surface density of small and large dust grains used in our fiducial model
  is shown in Fig.~\ref{density}. 

\begin{figure}
   \centering
   \includegraphics[angle=0,scale=1.0]{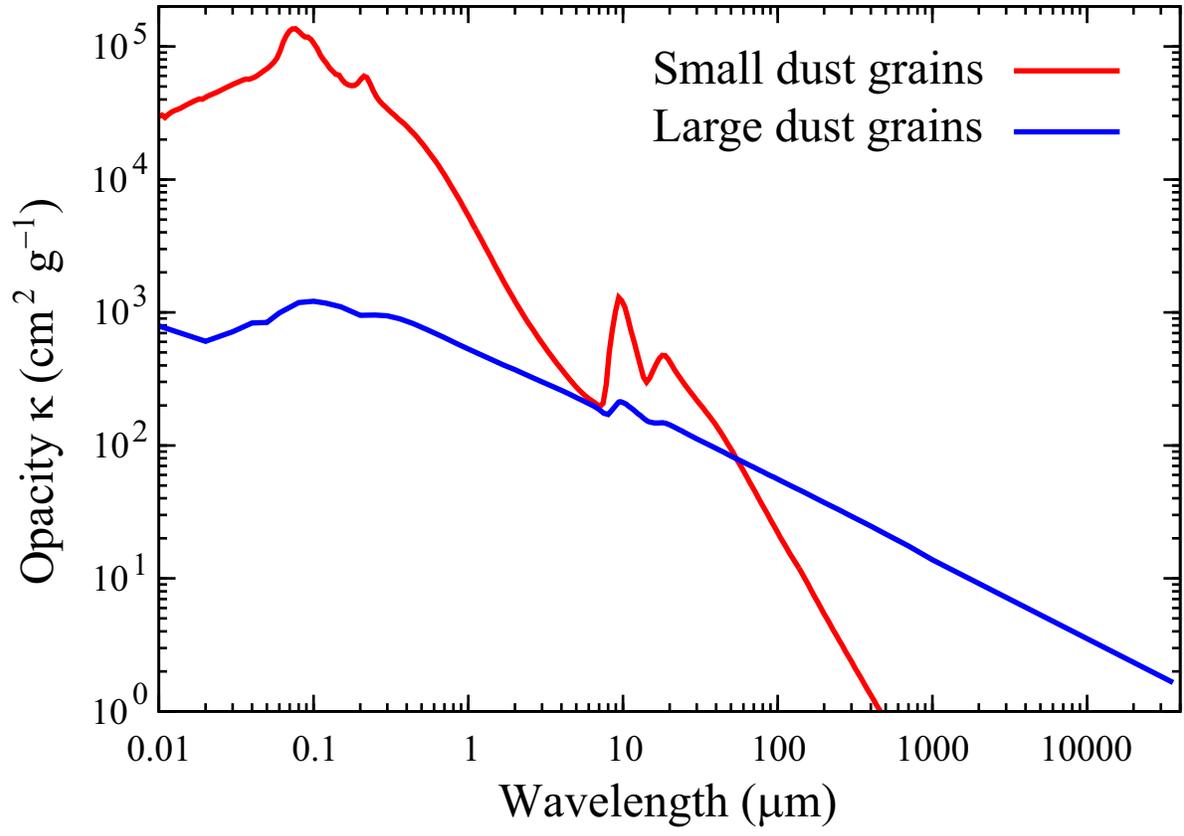}
   \caption{
     Opacity of small and large dust grains used in our modeling efforts.
   }\label{opacity}
 \end{figure}

  Dust properties used in this paper follow \citet{dong12b}:
  small dust grains from the standard interstellar-medium dust model (a composition of silicates and graphites; a size distribution of 
  $n(s) \propto s^{-3.5}$ from $s_{\rm min} =$0.0025~$\mu$m to $s_{\rm max} =$0.2~$\mu$m) in \citet{kim94} and large dust grains 
  (a composition of carbons and silicates; a size distribution of $n(s) \propto s^{-3.5}$ from $s_{\rm min} =$0.01~$\mu$m 
  to $s_{\rm max} =$1000~$\mu$m) from Model~2 in \citet{wood02}. Opacity ($\kappa$) of small and large dust grains used in
  our modeling efforts is shown in Fig.~\ref{opacity} \citep[see also Fig.~1 in][]{dong12b}.

  \subsection{Results of Modeling Efforts}\label{sec:result}
  
  We have synthesized the NIR polarimetric image, the SED, and the dust continuum of PDS~70 by varying the total mass of the disk 
  ($M_{\rm disk}$; assuming gas-to-dust mass ratio of 100), mass fraction ($f$) of large dust grains in the total mass of dust grains,
  scale height ($h$) at 100~AU, depletion factor ($\delta$), size of a gap in large dust grains ($R_{\rm gap}^{\rm l}$),
  and characteristic radius ($R_{\rm c}$). The size of a gap in small dust grains ($R_{\rm gap}^{\rm s}$) 
  is set to 65~AU because NIR imaging 
  constrains  
  this size \citep{hash12,dong12b}. 
  Most parameters 
  are similar to those in \citet{hash12} and \citet{dong12b} 
  except the depletion factor and the characteristic radius. Since the procedure in synthesizing NIR polarimetric images 
  in \citet{hash12} and \citet{dong12b} 
    did not subtract a
  `polarized halo \citep[described in \S~2.1 in][]{hash12}', the depletion factor and 
  the characteristic radius in our fiducial models are slightly different from our previous studies. To simplify, 
  we assume
  the depletion factor and the characteristic radius between small and large dust grains are same. In addition, 
  the scale height inside and outside the gap is also the same. The parameters in the fiducial model 
  and previous model 
  are shown in Table~\ref{param}. 

\begin{figure}
   \centering
   \includegraphics[angle=0,scale=.75]{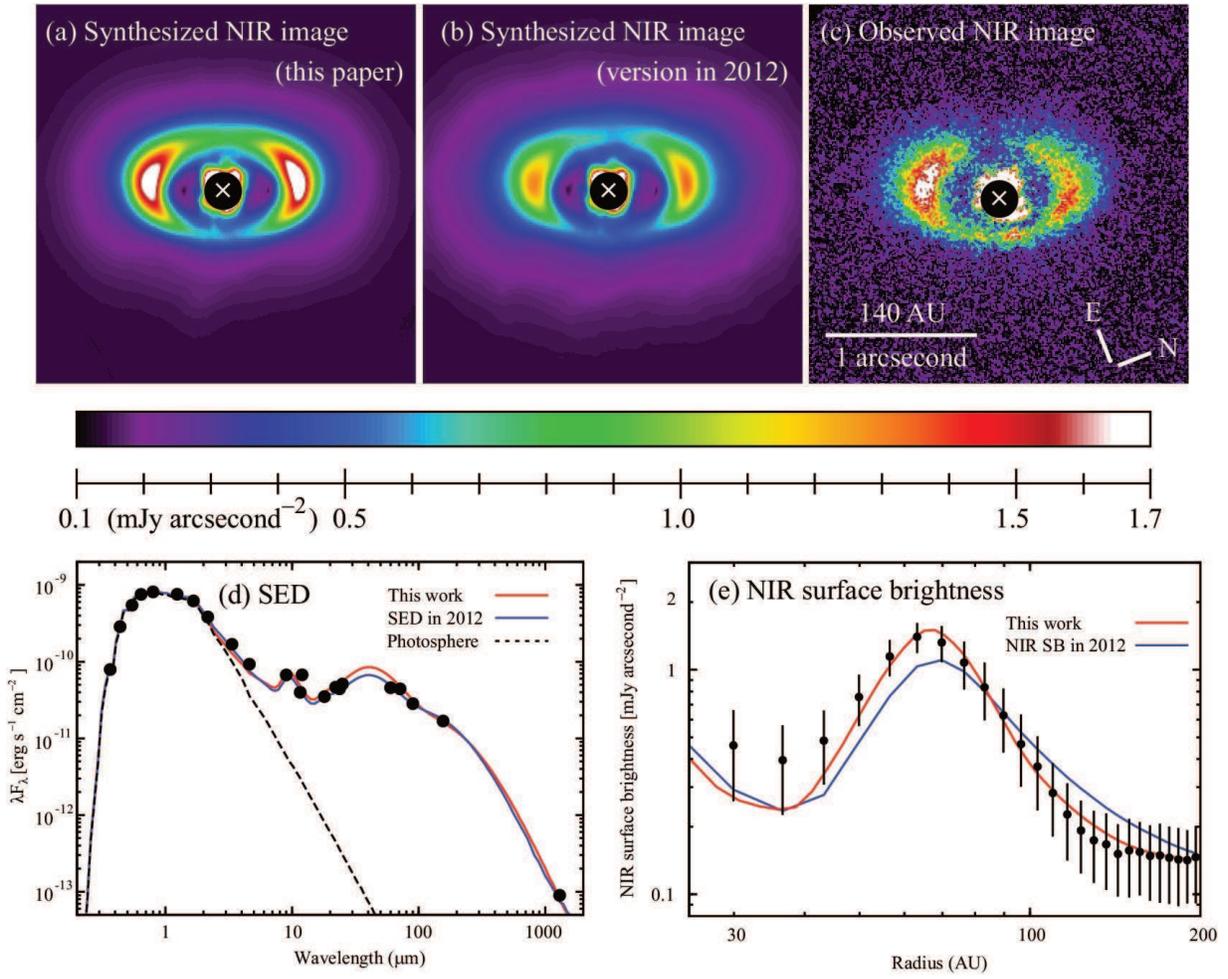}
   \caption{
     Comparisons of the synthesized NIR polarized intensity images in our fiducial model and previous model.
     (a) and (b): The synthesized images in our fiducial model and previous model, and (c): NIR observational results 
     in Fig.~\ref{res:dust_pds70}(c). Crosses represent the astrometric point
     \citep[14:08:10.125, $-$41:23:52.81;][]{cutr13}. The synthesized images in panel (a) and (b) are processed
     with subtracting the polarized halo and adding the offset value (see \S~\ref{sec:result}).
     (d): The synthesized SED and (e): radial profiles along the major axis in panel (a) and (b).
     Note that error bars in the SED are smaller than plots.
   }\label{comparison}
 \end{figure}

    Fig.~\ref{comparison} shows the difference of synthesized NIR images between the fiducial model and the previous model.  
    The SEDs in two models (Fig.~\ref{comparison}d) are consistent, 
    whereas the radial profiles along the major axis are slightly different (Fig.~\ref{comparison}e). After modeling,
    we subtracted the polarized halo from two synthesized images using IRAF\footnote{
  IRAF is distributed by the National Optical Astronomy Observatories, which are operated
  by the Association of Universities for Research in Astronomy,
  Inc., under cooperative agreement with the National Science Foundation.}
    as a same manner in \citet{hash12}.
    We also added an offset value of 0.14~mJy~arcsec$^{-2}$ to the synthesized NIR images.
    Since polarized intensity (PI) is calculated as square-root of sum of squares in stokes $Q$ and $U$,
    i.e., $PI = \sqrt{Q^{2} + U^{2}}$, the noise becomes always positive. The value of 0.14~mJy~arcsec$^{-2}$ is the averaged
    noise level in our NIR obsevations. These two operations, subtracting the polarized halo and adding the offset value of
    averaged noise, result in the difference of radial profiles of two models.

\begin{figure}
   \centering
   \includegraphics[angle=0,scale=.80]{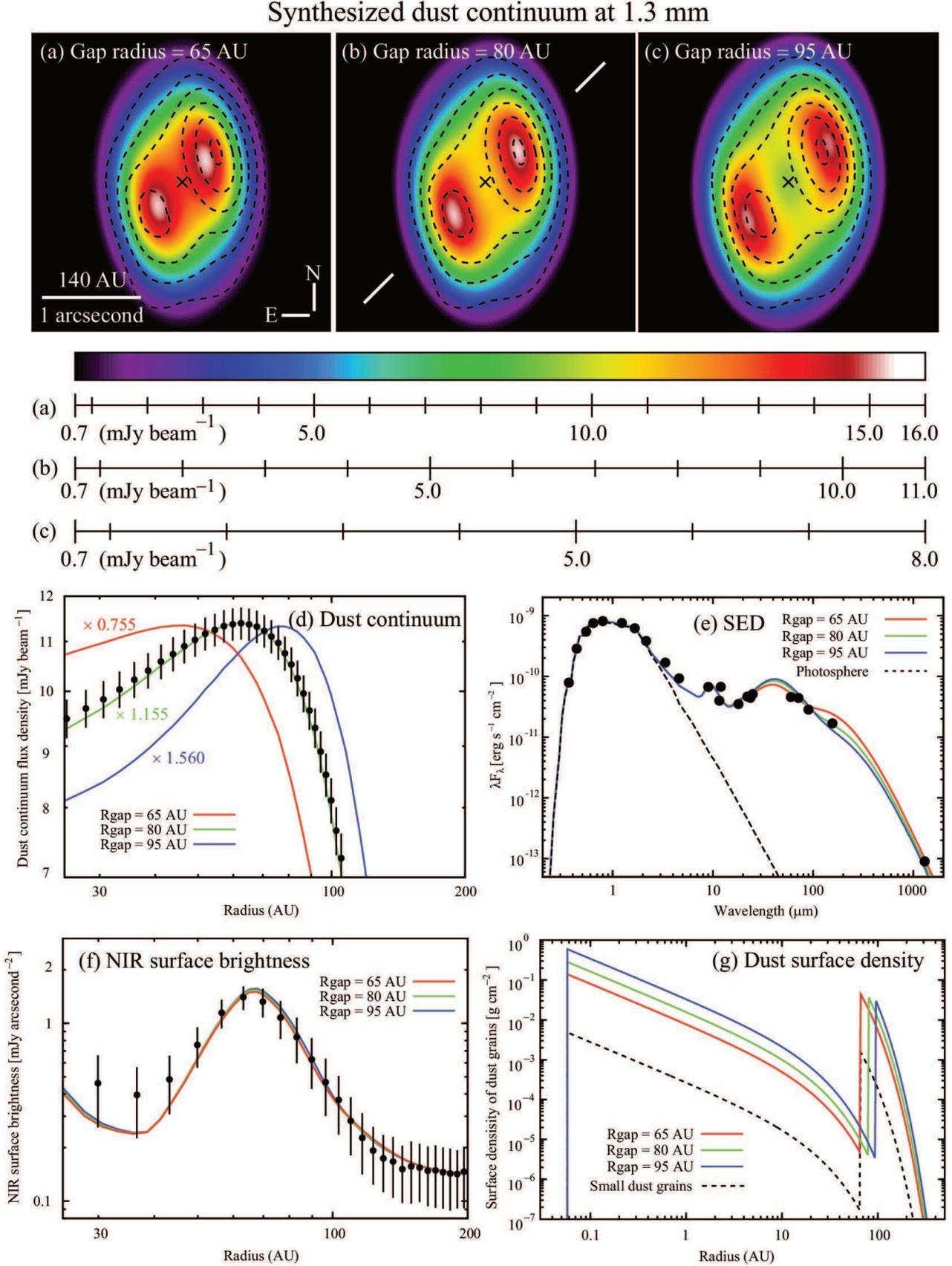}
   \caption{
     Our modeling efforts in synthesizing the dust continuum, the SED, and the NIR polarized-intensity surface brightness.
     (a) to (c): the synthesized dust continuum images varying a radius of a gap in large dust grains (65, 80, and 95~AU).
     A gap-radius in small dust grains is set to 65~AU and other parameters follow those in the fiducial model listed 
     in table~\ref{param}. Contours are taken from our SMA observations and same with Fig.~\ref{res:dust_pds70}(a).
     (d): The radial profile of the synthesized dust continuum at PA~$=$$135^{\circ}$ (along a white line in panel~b),
     (e): the synthesized SED, (f) the radial profile of the NIR polarized-intensity surface brightness at PA~$=$$160^{\circ}$, and
     (g): the surface density used in our modeling efforts. 
     Colored solid-lines indicate results of gap-radii in large dust grains of 65 (red), 80 (blue), and 95~AU (green).
     In panel (d), the flux densities of the synthesized dust continuum with gap-radii of 65, 80, and 95 are
     multiplied by 0.755, 1.155, and 1.560, respectively, for the presentation purpose.
     In panel (e), error bars in the SED are smaller than plots.
   }\label{res:model_pds70}
 \end{figure}

  As a result of our MCRT modeling with $9\times10^{8}$ photons, we found the apparent size of the gap is different 
  (Fig.~\ref{res:model_pds70}): 
  80 and 65~AU in a radius in large and small dust grains, respectively. We also show the dust continuum images with fiducial parameters
  in Table~\ref{param} 
  with different sizes
    of a gap in large dust grains ($R_{\rm gap}^{\rm l} =$65, 80, and 95~AU) 
  in Fig.~\ref{res:model_pds70}. Dust continuum images are convolved with a 2D elliptical Gaussian function of 
  $0\farcs97\times0\farcs48$ at PA~$=$~12.33$^{\circ}$ which is a beam-size in our SMA observations. 
  The synthesized dust continuum with a gap-radius of 80~AU (Fig.~\ref{res:model_pds70}b) is 
  matched well 
  with SMA observations
  except for the asymmetries due to axisymmetric structures assumed in our modeling efforts. 
  We compare the averaged radial profile of the 
  observed dust continuum with that of synthesized dust continuum (Fig.~\ref{res:model_pds70}d). The radial profiles along
  the apparent major-axis at PA of 135$^{\circ}$ are averaged in the northwest and southeast parts. Though the shapes of the profiles 
  are reproduced well
  with the synthesized dust continuum with a gap-radius of 80~AU, the flux density of the synthesized 
  dust continuum needs to be multiplied by 1.155 in order to obtain the consistent absolute flux density. This result, that
  the radial profile of the observations is brighter than those of the synthesized dust continuum, implies the flux density 
  may be gathered in the northwest rather than azimuthally symmetric.

  The synthesized SEDs in Fig.~\ref{res:model_pds70}(e) do not differ significantly except around 200~$\mu$m.
  Large dust grains in the disk-midplane are heated by the emission from the upper layer of the disk \citep[c.f., Fig.~1 in][]{dale06}.
  Since the temperature of the upper layer of the disk generally decreases with the distance 
  from
  a central star,  
  the disk-midplane irradiated by the upper layer also cools down with the radius. The different gap-radius in large dust grains
  would reflect the amount of thermal emission around 200~$\mu$m in the SED. The three synthesized NIR polarimetric images are reasonably
  consistent with each other (Fig.~\ref{res:model_pds70}f). Since the opacity at NIR wavelengths
  in small dust grains is $\sim$10 times larger than that in large dust grains (Fig.~\ref{opacity}); and the disk in large dust grains
  is located in the midplane, minor geometric differences of the disk in large dust grains do not affect the NIR scattering image.

    Our modeling efforts in this paper are relatively simple, and the above fiducial model would not be 
    the best solution.
      Even so, our disk model with a gap consistently explains the the observed images at NIR and millimeter
      wavelengths and the SEDs. In particular, our modeling suggests different gap radii for small and large grains.
    Further complicated disk models (e.g., not sharp gap-edge) and fitting the visibility profile of the dust continuum 
    will be presented elsewhere. 

  \section{Discussion}\label{sec:discuss} 

  We discuss scenarios to account for the observational features of PDS~70, especially the asymmetries and the different 
  gap-radii. Before discussing the possible origin of the complicated disk around PDS~70, we summarize six main observational 
  properties of PDS~70.
  \begin{itemize}
  \item[(a)] Infrared to millimeter observations and our modeling suggest that 
    the depletion of the gas in the disk as well as large and small dust grains within tens of AU in radius 
    (\S~\ref{sec:obs} and \ref{sec:result}).
  \item[(b)] The contrast ratio in the dust continuum of 1.4 (\S~\ref{sec:obs}).
  \item[(c)] The above model suggests 
    different gap-radii in small and large dust grains, 65 and 80~AU, respectively (\S~\ref{sec:result}). 
  \item[(d)] The existence of the optically thick inner disk of small dust grains at $\sim$0.1~AU \citep{hash12}.
  \item[(e)] 
    The upper limit of a planetary mass in the gap is estimated to be $\sim$2~$M_{\rm Jup}$ in our previous studies \citep{hash12}  
    based on a hot-start planet model \citep{bara03} and the lack of detection of any point sources in the image.
  \item[(f)] The low mass accretion; PDS~70 is a weak-lined T Tauri star (a H$\alpha$ equivalent width
  of 2.0~\AA; \citealt{greg02})
  \end{itemize}
  Thereby, scenarios proposed to explain both the asymmetries and the different gap-radii of PDS~70 should also explain 
  all above observational 
  constraints. In this section, we firstly discuss the possible origin of the asymmetries, and then, different gap-sizes independently. 
  Finally, we discuss the possible scenario responsible for all observational features and suggest how to 
  further constrain
  a proposed scenario in future observations. 

  \subsection{Azimuthal Asymmetry}\label{sec:morphology}
 
  Fig.~\ref{res:dust_pds70} show an asymmetric brightness in the dust continuum at 1.3~mm and NIR polarized intensity. 
  Such asymmetries in the dust continuum have been detected in other transitional disks, e.g., LkH$\alpha$~330 
  \citep {brow08} and SAO~206462 \citep{brow09}. The contrast ratio of the peak-emission to the opposite-side of the disk is likely to 
  be higher among transitional disks around intermediate mass stars (e.g., $\sim$30 in HD~142527; \citealt{casa13,fuka13} and $\sim$130 
  in WLY~2-48; \citealt{vand13}) than among solar-mass stars (e.g., $\sim$1.1-1.2 in UX~Tau~A, DoAr~44, and LkCa~15; 
  \citealt{andr11}). PDS~70's contrast ratio of 1.4 is similar to 1.3-1.5 in SAO~206462 \citep{brow09,pere14} and slightly higher 
  than that among solar-mass stars. The physical interpretation of the asymmetries is still under debate: gravitational interaction 
  between 
  planet and disk \citep[e.g.,][]{kley12}, large anticyclonic vortex induced by the Rossby wave instability \citep[e.g.,][]{rega12}, 
  eccentric disk formed by a massive planet \citep[e.g.,][]{kley06}, and an azimuthal variance of temperature and opacity. 
  In this sub-section, we discuss the potential origin of asymmetric features observed in PDS~70.
   
  {\it Disk-planet interaction} --- An embedded 
  massive planet with a mass of $\gtrsim$1~$M_{\rm Jup}$ could open a gap and excite a spiral density wave in the disk due to tidal 
  interactions, while a less massive planet could only excite a wake 
  because the perturbation is not strong enough (see \citealt{papa07}; \citealt{kley12} for a review). 
  These interactions could also 
  lead to exchange of angular momentum of the gas and dust grains 
  in the disk, and are expected to result in asymmetric structures in the disk as well. Actually, previous studies have demonstrated 
  the asymmetries in the surface density of the disk using hydrodynamic simulations \citep{dods11,isel13}.
  Though the relationship between disk-planet interaction and asymmetric structures in the disk still remains to be intensively 
  investigated, a single massive planet could tend to induce azimuthally symmetric structures except spirals \citep{kley12},
  whereas multiple massive planets could create azimuthal asymmetries \citep{dods11,isel13}.
  Hence, the existence of multiple planets would be one of the key points to explain the observed asymmetries in PDS~70.
  The possible existence of (unseen) multiple planets opening the gap around PDS~70 has already been argued in \citet{hash12,dong12b}
  and been supported by following two aspects: the large gas-gap
  around PDS~70 with a radius of $\sim$65~AU (Fig.~\ref{res:dust_pds70}c) 
  could be explained by only multiple planets \citep{zhu11,dods11}; a gap structure (optically thick inner and outer disks separated 
  by an optically thin region) inferred from NIR imaging (Fig.~\ref{res:dust_pds70}c) and the SED 
  (Fig.~\ref{res:model_pds70}e) is most like due to disk-planet interaction rather than other inside-out scenarios (grain growth; 
  photoevaporation; see \S~\ref{sec:intro}). Furthermore, \citet{hash12} put an upper limit of the planetary mass of only 
  $\sim$2~$M_{\rm Jup}$ in the gap based on $L'$ band (3.8~$\mu$m) imaging, 
thus being unable to discard the existence
  of unseen massive planets with a mass of $\sim$1~$M_{\rm Jup}$. 
  Therefore, though further quantitative comparisons between observations and numerical simulations (e.g., comparisons 
  between the observed PDS~70's contrast ratio of 1.4 and simulations) are still necessary, we anticipate the observed asymmetries 
  in the disk of PDS~70 might be induced by interactions between unseen multiple planets and the disk.

  {\it Anticyclonic vortex} --- 
  Asymmetric structures in a disk can be explained without any planets. In the last decade, 
  several transitional disks show asymmetric dust continuum emission 
  \citep{piet05,brow09,andr11,pere14} and has been explained with large anticyclonic vortex potentially formed 
  by the Rossby wave instability \citep[RWI;][]{love99,rega12}. RWI is excited at steep gas-pressure gradients 
  in an overdense region where is 
  thought to be an outer edge of the `dead zone' \citep{gamm96}. 
  The gas in the disk is generally transported from the outer disk through 
  mass accretion, possibly triggered by the magnetorotational instability \citep[MRI;][]{balb91}. MRI becomes active when the ionization 
  state of the gas is sufficient, meaning an ionized upper layer of a disk and a less dense outer-disk are prone to induce MRI. 
  Conversely, a dense region in 
  the disk-midplane would have less effective ionization because of gas self-shielding against ionization radiation, resulting in a
  dead zone where mass accretion by MRI is strongly reduced. As a result, a gas-overdense region is expected to form at the boundaries of
  the dead zone \citep{varn06,terq08}, leading RWI to launch anticyclonic vortices. The radius of the outer edge in the dead zone 
  strongly depends on the amount of dust grains \citep{sano00} because degrees of ionization in the gas are suppressed by 
  recombination of ions and electrons at grain surfaces. Assuming the minimum-mass solar nebula \citep{haya85}, \citet{sano00} estimated
  the dead zone's outer edge is about 20~AU in radius. Since this radius is also a function of the disk's properties 
  \citep[e.g., a vertical magnetic field, density, and a temperature of a disk;][]{okuz11}, anticyclonic vortices might take place at 
  a radius of 80~AU in the case of PDS~70. However, this scenario would not explain the gap structure even if it explains
  asymmetric structures; 
    hence it is unlikely that an anticyclonic vortex induced at the outer-edge of the dead zone could be the main scenario.

  Steep gas-pressure gradients can also form at the gap-edge sculpted by massive planets \citep{papa07,kley12}, 
  resulting in anticyclonic vortices \citep[e.g.,][]{li05,lin10,zhu14a,zhu14b}.
  Coriolis force in a vortex effectively leads large dust grains into the central 
  vortex \citep[e.g., Fig.~1 in][]{barg95}, resulting in an increase in the surface density of the large dust grains 
  by more than a factor of 10-100 for fastest drifting particles \citep{zhu14a,zhu14b}. 
    \citet{vand13} modeled the asymmetric structures in the dust continuum around WLY~2-48, which has the large contrast ratio of 130, 
    with a vortex-shaped dust trap triggered by a companion. 
    Although PDS~70 has a contrast ratio in flux density of only 1.4,
    this value may be due to the convolution of the large beam size ($\sim$1$''$) in our observations. Thus
    this scenario of the vortex 
    at the planet-induced gap-edges would be applied to explain not only the large contrast ratio in HD~142527 and WLY~2-48 
    with $\sim$30 and $\sim$130, respectively, but also the case of PDS~70.
  
  {\it Variance of Temperature and Opacity} ---
  When a disk is optically thin, the observed azimuthal flux density $F_{\nu}$ is
  \begin{eqnarray}
    F_{\nu} \approx \kappa_{\nu} \times M_{d} \times B_{\nu}(T_{d}) \times d^{-2}, 
  \end{eqnarray}
  where $\kappa_{\nu}$ and $T_{d}$ are the opacity and temperature of a disk with a mass of $M_{d}$, 
  $B_{\nu}$ is the blackbody intensity, and $d$ is the distance \citep[c.f.,][]{drai06}. Since  
  the Rayleigh-Jeans limit ($h\nu \ll kT_{d}$) can be applied at (sub-)millimeter wavelengths \citep{beck90,beck91},
  $F_{\nu}$ is 
  \begin{eqnarray}
    F_{\nu} \propto \nu^{2} \times \kappa_{\nu} \times M_{d} \times T_{d}.
    \label{flux}
  \end{eqnarray}
  Hence, assuming the azimuthally-symmetric dust opacity $\kappa_{\nu}$ and mass-distributions of the dust grains, azimuthal 
  distributions of the observed flux density depend on disk temperature. As mentioned in \S~\ref{sec:result}, since 
  the disk-midplane is heated by the emission from the upper layer of the disk, if the stellar radiation is interrupted by an asymmetric 
  inner disk (e.g., a warp; an asymmetric puff-up), a shadow cast from the inner disk might cool down disk temperature in the outer 
  disk. 
  Disk temperature of large dust grains in the mid-plane around $r=$80~AU of PDS~70 
  simulated in our modeling efforts is 23~K, corresponding
  to the azimuthal temperature variance of 16-23~K using the contrast ratio of the flux density in PDS~70 with 1.4. 
  Thus, we cannot rule out this scenario based on our observations. 
    Though whether this scenario of shadowing can directly affect on (sub-)millimeter radiation and azimuthal redistribution
    in a (closer to) optically thin environment is still ambiguous, monitoring observations would be useful to test 
    the shadowing-scenario, because a shadow cast from the inner disk is co-rotating with the inner disk.
  The Keplerian orbital velocity ($\omega$) of the inner disk at 0.1~AU (where is optically thick in Fig.~\ref{density}) is given by 
  \begin{eqnarray}
  \omega \approx 27 \left( \frac{M}{0.8~{\rm M}_{\odot}}\right)^{\frac{1}{2}} \left( \frac{r_{{\rm d}}}{0.1~{\rm AU}} \right)^{-\frac{3}{2}} \ {\rm [deg~day^{-1}]},  
  \end{eqnarray}
  where $M$ is the mass of the central star and $r_{{\rm d}}$ is the orbital radius of the inner disk. Such rotation could be 
  observed within a few months. 
  Furthermore, a non-axisymmetric illumination from the central star \citep{taka14}
  possibly causes asymmetric distributions of the disk temperature. Assuming PDS~70 has the rotational velocity of $\sim$10~km~s$^{-1}$ 
  and a radius of 1.39~$R_{\odot}$ \citep{greg02}, the rotation period is $\sim$50~days. Such a periodic fluctuation
  could be also observable within a few months. 

  Note that though above discussions are based on the azimuthally-symmetric dust opacity, 
  asymmetrically distributed dust opacity could also affect the flux density. If the dust opacity $\kappa_{\nu}$ 
  is assumed to have a power-law dependence on frequency, $\kappa_{\nu} \propto \nu^{\beta}$ \citep{beck90,beck91}, 
  the observed azimuthal flux density in equation~(\ref{flux}) becomes $F_{\nu} \propto \nu^{\alpha}$ 
  (where $\alpha = \beta + 2$) assuming azimuthally-symmetric disk temperature and mass distributions of the dust grains. A power-law 
  index $\beta$ is known to strongly depend on size distributions and maximum sizes of dust grains \citep{drai06}. For instance, 
  interstellar medium (ISM) dust mainly consisting of sub-$\mu$m size grains which are less (sub-)millimeter emitters shows 
  $\beta \approx 1.7$ \citep[e.g.,][]{li01}, while circumstellar disks consisting of both small and large dust grains show 
  $\beta \sim 0$ \citep[e.g.,][]{andr07}. Thus, if grain growth locally takes place in the disk due to, for example, anticyclonic 
  vortices, asymmetric dust continuum emission could be observed. The parameter $\beta$ in $\kappa_{\nu}$ 
  of large dust grains in our modeling is 
  $\sim$0.6 (Fig.\ref{opacity}), and $\beta$ changing with $\delta \beta \sim$0.017 
  reproduces the contrast ratio of 1.4 of PDS~70 at 1.3~mm.
  When dust grains grow into larger dust grains in the disk, this chage in $\beta$  
  would be reasonable. Hence, though we 
  cannot exclude this scenario based on our single-wavelength observations, future high-resolution multi-color radio 
  interferometric observations with ALMA 
  could reveal azimuthal $\beta$ distributions, allowing the investigation of whether dust grains locally evolve in the disk.

  {\it Eccentricity} ---
  Numerical calculations expect that when a massive planet with a stellar-to-planet mass ratio of $q \gtrsim 3$-$5\times10^{-3}$ is 
  embedded in the disk, an eccentric disk can form at the gap-edge 
  \citep[eccentricity of up to $\sim$0.25;][]{kley06,atai13,rega14}. 
  Keplerian orbital velocity at the 
  apocenter of the elliptic orbits is slower than at the pericenter, resulting in a possible enhancement of disk materials at the 
  apocenter of the disk. This scenario makes a `traffic jam' instead of a dust-trap, suggesting a lower azimuthal contrast than 
  that in the vortex. According to \citet{atai13}, the contrast ratio of the disk mass between the apocenter and the pericenter is 
  $\sim$1.5 \citep[see Fig.~3 in][]{atai13}. This value is consistent 
  with that of PDS~70. However, the necessary 
  minimum-planetary-mass to induce an eccentric disk around PDS~70 would be $\sim$2-4~$M_{\rm Jup}$; such planetary-mass objects
  in the gap should be detected in our previous observations \citep[see Fig.~2f in][]{hash12}.
  One of the differences in this scenario compared to the above is 
  that the asymmetries in an eccentric disk are fixed at the same position instead of co-moving with planets or 
  vortices; hence future follow-up observations could confirm whether an eccentric disk forms around PDS~70.

  \subsection{Different Radial Structures in Dust Grains}\label{sec:structure} 
  
  Our observations and modelling efforts suggest that 
  the disk around PDS~70 has different radial structures in large and small dust grains: a gap of $r=$80~AU for large dust 
  grains and 65~AU for small dust grains (\S~\ref{sec:result}; Fig.~\ref{res:model_pds70}). In \S~\ref{sec:intro}, we mentioned 
  grain growth and photoevaporation as potential scenarios to explain the properties of transitional disks \citep[e.g.,][]{clar01,dull05}.
  As discussed by other authors \citep[e.g.,][]{espa14}, these two scenarios are expected to result in inside-out disk-dispersal.
  Since PDS~70 has an optically thick inner disk at $\sim$0.1~AU (Fig.~\ref{density}; \citealt{hash12}), these two would be less 
  dominant in PDS~70. Another possible scenario is disk-planet interaction \citep[e.g.,][]{kley12}. 
  However, {\it only} gravitational interaction between the disk and planet might result in similar radial structures in
  three disk components (the gas and the large and small dust grains), i.e., no radial difference; thus
  {\it only} disk-planet interaction would also be less likely in PDS~70.
  Other scenarios mentioned in \S~\ref{sec:intro} (dust filtration and radiation pressure) need planet(s) in the gap.
  In this sub-section, we briefly review and discuss the combinations of scenarios: disk-planet interaction with dust filtration 
  and/or radiation pressure.

  Dust filtration \citep{rice06,zhu12,pini12,deju13,owen14} is induced at the outer edge of the gap. Dust filtration has been originally 
  suggested to explain apparently contradictory transitional disk properties: a T-Tauri-star-like moderate mass accretion rate 
  ($\dot{M} \sim$10$^{-8}$ to 10$^{-9}$~$M_{\odot}$~yr$^{-1}$; \citealt{naji07}) with an optically thin or empty inner hole. 
  Since the disk-materials in the inner disk are resupplied from the outer disk through mass accretion, moderate mass accretion and 
  an inner hole are apparently inconsistent. To interpret these transitional disk properties, 
  theorists make use of a physical consequence that dust particles are trapped at gas-pressure maxima. When a dust particle embeds
  in a gaseous disk where the pressure has a radially negative gradient, since the gas rotates with the sub-Keplerian velocity,
  a dust particle feels a headwind and loses angular momentum. On the other hand, when a gaseous disk's pressure has a radially positive
  gradient, for example, at the outer edge of the gap, since the gas rotates with the super-Keplerian velocity, a dust particle feels 
  a tailwind and gains angular momentum. Eventually, dust particles are led into gas-pressure maxima. The gas can still accrete onto the 
  central star through the gap \citep{lubo06} even there are multiple planets in the gap \citep{dods11,zhu11}. 
  That is how large dust grains are distributed at gas-pressure maxima, whereas moderate mass accretion is still observable. 

    \citet{zhu12} demonstrated that though dust filtration with a single or multiple planets affects 
    (sub-)millimeter dust particles, which possibly accounts for (sub-)millimeter observations of large
    gapped ($r \gtrsim$20~AU) transitional disks with moderate mass accretion of $\gtrsim$10$^{-8}$ to 10$^{-9}$~$M_{\odot}$~yr$^{-1}$,
    the filtration process hardly affects (sub-)micron sized dust particles since small dust 
    grains are generally coupled well with the gas.
    This can nicely explain the missing gaps in NIR scattered light images of some transitional disks \citep{dong12a},
    such as SR~21 \citep{foll13}, SAO~206462 \citep{muto12}, and MWC~758 \citep{grad13}.
  Subsequent studies in dust filtration \citep{pini12} calculated the position of gas pressure maxima depends
  on disk viscosity, the planetary mass, and location (e.g., a gas pressure maximum at 30 and 50~AU when 
  1 and 9~$M_{\rm Jup}$ at 20~AU, respectively, 
  in Fig.~1 in \citealt{pini12}); \citet{deju13} showed observed radii of the gap in the NIR and sub-millimeter wavelengths could also 
  depend on the planetary mass and location (e.g, gap radii in the NIR and sub-millimeter wavelengths are 20 and 30~AU, respectively,
  when a planet with 1~$M_{\rm Jup}$ at 20~AU in Fig.~2 in \citealt{deju13}). The radius of the gap in the NIR wavelengths could reflect
  the scattered position at the outer gap-edge of the disk in small dust grains carved by a planetary perturbation, 
  while the radius in the 
  (sub)-millimeter wavelength reflects the peak position of the dust surface-density (i.e., gas pressure-maximum which slightly locates 
  at the far side; see Fig.~7 in \citealt{deju13}). 
  The dependence of the dust-gas coupling on the size of the dust grains leads
  to different gap-radii at different wavelength, as seen in PDS~70.
  Such disk systems have been reported in SAO~206462 \citep{garu13,pere14}, HD~100546 \citep{aven14,pine14}, 
  and Sz~91 \citep{tsuk14}; thus PDS~70 is not unique.
  
  As pointed out by \citet{zhu12}, one problem in dust filtration is that it is difficult 
  to account for the low NIR excess generally seen in transitional disks.
  In addition to dust filtration, to reduce the NIR excess generated from small dust grains, `a planetary accretion luminosity' 
  was proposed by \citet{owen14} in which small dust grains are expected to push back by radiation pressure from an accreting planet, 
  resulting in allowing the dust-free gas to accrete onto the central star \citep{owen14}. This scenario would be useful to explain
  the properties of general transitional disks with the reduced NIR excess and moderate mass accretion. However,
  in contrast to general transitional disks, weak-line T~Tauri star PDS~70 should have low mass accretion (a H$\alpha$ equivalent width
  of 2.0~\AA; \citealt{greg02}). 

  \subsection{Possible Origin of Complicated Disk-Structures}\label{sec:origin} 

  Though we have discussed the possible scenarios to explain the asymmetries in PDS~70, all proposed scenarios 
  except eccentricity are hardly excluded based on our single epoch and single wavelength observations with SMA 
  (\S~\ref{sec:morphology}).
  On the other hand, we found disk-planet interaction $+$ dust filtration could account for different radial structures
  in small and large grains (\S~\ref{sec:structure}).
  In this sub-section, we discuss whether disk-planet interaction $+$ dust filtration is potentially responsible for
  the complicated structures of the disk around PDS~70.

  Planet(s) in the gap could reduce mass accretion and might explain the low mass accretion of PDS~70. 
  Though mass 
  accretion onto a planet is not conclusively understood, in general, a more massive planet (e.g., 10~$M_{\rm Jup}$) 
  or multiple planets (e.g., 1~$M_{\rm Jup}$) could reduce mass accretion onto the central star. 
  Since the existence of such a single 10~$M_{\rm Jup}$ planet was already ruled out in our previous studies \citep{hash12},
  accreting multiple planets remain as a possible scenario. 
  If a single planet with $\sim$1~$M_{\rm Jup}$ can be assumed to have sufficient mass accretion onto the planet itself
  and induce dust filtration, 
  this planet might be responsible for the large-gapped disk-structure in the dust continuum \citep{pini12}
  and low accretion rate of PDS~70. 
  However, a single planet would likely induce a relatively symmetric structure in the disk \citep{kley12} and thus be 
  inconsistent with the asymmetries observed in PDS~70. Meanwhile, a single planet with $\sim$1~$M_{\rm Jup}$ 
  could create the asymmetries in the disk if vortices take place at the gap-edge sculpted by this single planet. 
  However, a single planet is expected to create a narrow gap in the gas and small dust grains 
  \citep[i.e., $r\lesssim$15~AU;][]{zhu11,dods11}; 
  hence a single accreting planet inducing vortices would be ruled out as the origin of the large NIR gap around PDS~70. 
  Remaining accreting multiple planets could explain both of low mass accretion and the asymmetries of the large gapped-disk in PDS~70. 
  These combinations could also induce dust filtration at the gap-edge and are likely to account for 
  the different gap-radii in the NIR and millimeter wavelengths.

  To summarize, 
  accreting multiple planets inducing dust filtration at the outer gap-edge could explain all observed properties 
  of PDS~70: different observed radii of the gap in the different wavelengths; low mass accretion onto the central star; 
  a depletion of disk components 
  inside the gap except the optically think inner disk; no very massive planets in the gap. However, direct observational evidence 
  of dust filtration has not been detected in any disk system. One conclusive observational proof of dust filtration
  would be `a gap in the gas' 
  \citep[see, e.g., Fig.~7 in][]{deju13}. As mentioned, in the outside of the planetary orbit, a combination of positive and negative 
  gas-pressure gradients is critical to trap large dust grains at gas-pressure maxima. Inside the planetary orbit, gas-pressure would 
  have only negative gradients, resulting in the depletion of large dust grains. Future ALMA observations with high-resolution 
  and high-sensitivity will enable the detection of a gas gap. Furthermore, detecting any planets in the gap 
  would be important to verify whether
  planets form gap structures in the dust grains and induce dust filtration. A hint to enhance the detectability of planets in the 
  optical/NIR wavelength may be the planetary accretion luminosity \citep{owen14}. Though a low-mass binary, a companion in the 
  gap of HD~142527 (a mass of 2.2~$M_{\odot}$; a distance of 145~pc; an age of 5~Myr; \citealt{verh11}) exhibit 
  an accretion luminosity and 
  was detected in the narrow band of H$\alpha$ \citep{clos14}. Therefore, integral field spectroscopy combined with angular differential
  imaging \citep[ADI;][]{maro06} or spectral differential image \citep[SDI;][]{raci99} with ADI focusing on accretion-sensitive lines
  such as H$\alpha$ (0.656~$\mu$m); Pa$\beta$ (1.282~$\mu$m); Br$\gamma$ (2.166~$\mu$m); H$_{2}$~$v$$=$$1-0$~S(1) (2.122~$\mu$m)
  may increase the detectability and allow discovery of accreting planets in the gap.

  \section{Conclusion}\label{sec:conclusion} 

  We present the dust continuum at 1.3~mm and $^{12}$CO~$J=2\rightarrow1$ line emission of the pre-transitional disk
  around PDS~70 obtained with SMA, and compare these with NIR (1.6~$\mu$m) polarized intensity image from our previous studies
  to investigate the radial distributions of the gas and the large (millimeter size) and small ($\mu$m size) dust grains
  in the disk. The main observational results are as follows.
  \begin{itemize}
  \item Three disk-components (the gas and the large and small dust grains) show the deficit of flux density 
    in the central region 
    of the disk, suggesting depletion of the materials except small dust grains.
    High-resolution observations using 
ALMA would give more stringent constraints on the gas and large dust grains in the inner disk.
  \item 
Both large and small dust grains show azimuthal asymmetries. 
    The contrast ratio of the flux
    density at the peak position to that at the opposite side is 1.4 in the dust continuum. Since other transitional disks around
    solar-mass stars have the contrast ratio is $\sim$1.1-1.2, the PDS~70's contrast ratio is slightly higher among solar-mass stars.
  \item 
    The results of Monte Carlo radiative transfer models suggest different radii
    for the gap edges: 80 and 65~AU for large and small grains, respectively.
  \item We observed a velocity gradient from the northwest to southeast in $^{12}$CO~$J=2\rightarrow1$ line emission, suggesting 
    a rotating disk around PDS~70. The position-velocity  diagram of the CO gas infers the dynamical mass of PDS~70 of 
    0.6-0.8~$M_{\odot}$, which is consistent with the spectral type of K5.
  \end{itemize}

  We have discussed the possible origin of the azimuthal asymmetries in the disk. Though definite conclusions are not reached from
  our SMA and previous Subaru observations, we anticipate gravitational interaction between multiple planets and the disk, or
  azimuthal temperature and opacity variance, or anticyclonic vortex
  could be responsible for the observed asymmetries. 
  Future follow-up observations to test opacity distributions and 
  azimuthal rotations of the asymmetries are expected to put further constraints.

  Finally, gap formation with different gap-radii in large and small grains around PDS~70 have also been discussed. As discussed in
  \citet{hash12} and \citet{dong12b}, 
  both grain growth and photoevaporation are unlikely because they require an inside-out process
  that directly contradicts the presence of optically thick inner disk around PDS~70.
  We argue that
  combinations of accreting multiple planets and dust filtration explain the observations of 
  the gap structure, low mass accretion, and different radii in the outer gap-edge as well as 
  the asymmetries of PDS~70. Future observations to confirm the existence of multiple planets and dust filtration are desirable.
  In ALMA cycle~2, a spatial resolution will be comparable to 0.$''$1-0.$''$2 of NIR observations.
  Such a higher spatial resolution could resolve local density-enhancement possibly due to anticyclonic vortices at the gap-edge;
  thus our understanding of complicated disk-structures around PDS~70 will be improved in the near future.

  \bigskip  

  We are grateful to an anonymous referee for providing
  useful comments leading to an improved paper.
  We appreciate support from the SMA staff.
  This work is partly supported by a Grant-in-Aid for Science Research in
  a Priority Area from MEXT Japan, by the Mitsubishi Foundation, 
  and by the U.S. National Science Foundation under Awards No. 1009203 and 1009314.
  This work is partially supported by Grant-in-Aid for JSPS Fellows (No. 25-8826).

\clearpage

\begin{deluxetable}{lccccc}
\tablecolumns{6}
\tablewidth{0pt} 
\tabletypesize{\normalsize}
\tablecaption{\label{table:pds70obs}PDS~70 Observations}
\tablehead{
\colhead{Date} & \colhead{Configuration} & \colhead{Number of} & \colhead{Beam} & \colhead{SSB T$_{sys}$} & \colhead{LO frequency}\\
\colhead{}     & \colhead{}              & \colhead{Antennas}  & \colhead{Size} & \colhead{(K)}          & \colhead{(GHz)}
} 
\startdata
2012 Dec 13 & Compact       & 7 & $3\farcs0\times6\farcs3$ & 170-200 & 225.507 \\
2013 Apr 15 & Compact       & 7 & $5\farcs7\times3\farcs0$ & 170-350 & 225.501 \\
2012 Dec 27 & Extended      & 7 & $1\farcs8\times1\farcs0$ & 230-360 & 225.227 \\
2013 Jan 03 & Extended      & 7 & $1\farcs8\times1\farcs0$ & 140-200 & 225.230 \\
2013 Feb 04 & Very extended & 8 & $0\farcs9\times0\farcs4$ &  85-120 & 224.859
\enddata
\end{deluxetable}

\clearpage

\begin{deluxetable}{ccccccccccc}
\tablecolumns{6}
\tablewidth{0pt} 
\tabletypesize{\normalsize}
\tablecaption{\label{param}Parameters in our fiducial model and previous model}
\tablehead{
\colhead{(Column 1)} & \colhead{(2)} & \colhead{(3)} & \colhead{(4)} & \colhead{(5)} & \colhead{(6)} & 
\colhead{(7)} & \colhead{(8)} & \colhead{(9)} & \colhead{(10)} & \colhead{(11)} \\
\colhead{Model} &
\colhead{$M_{\rm disk}$} & \colhead{$f$} & \colhead{$R_{\rm c}$} & \colhead{$R^{\rm l}_{\rm gap}$} & \colhead{$R^{\rm s}_{\rm gap}$} & 
\colhead{$h^{\rm l}_{\rm 100AU}$} & \colhead{$h^{\rm s}_{\rm 100AU}$} & \colhead{$p$} & \colhead{$q$} & \colhead{$\delta$} \\
\colhead{} & \colhead{($M_{\rm Jup}$)} & \colhead{} & \colhead{(AU)} & \colhead{(AU)} & \colhead{(AU)} &
\colhead{(AU)} & \colhead{(AU)} & \colhead{} & \colhead{} & \colhead{}
} 
\startdata
This work     & 4.5 & 0.9667 & 20 & 80 & 65 & 2.00 &  8.00 & 1.25 & 1.00 & $10^{-4}$ \\
Previous work & 3.0 & 0.9667 & 50 & 65 & 65 & 2.00 & 10.00 & 1.20 & 1.00 & $10^{-3}$ 
\enddata
\tablecomments{
Column 2: total mass of the disk (assuming a gas-to-dust ratio 100). 
Column 3: mass fraction of big dust in total dust. 
Column 4: characteristic radius in equation~\ref{eq_sig}.
Column 5 and 6: gap-radius of the disk in large and small dust grains. 
Superscripts `l' and `s' indicate large and small dust grains, respectively.
Column 7 and 8: scale height at 100 AU. 
Column 9: power index $p$ in the scale height $h \propto R^{p}$.
Column 10: power index in the surface density in equation~\ref{eq_sig}.
Column 11: depletion factor of the large and small dust disk.
}
\end{deluxetable}

\clearpage

\clearpage

\clearpage

\clearpage

\clearpage

\clearpage

\clearpage

\begin{thebibliography}{}
  %  \bibitem[Allen \& Cragg (1983)]{alle83} Allen,~D.~A. \& Cragg,~T.~A. 1983, MNRAS, 203, 777  
  \bibitem[Alexander, Clarke, \& Pringle(2006)]{alex06} Alexander,~R.~D., Clarke,~C.~J., \& Pringle,~J.~E. 2006, \mnras, 369, 229
  \bibitem[Alexander \& Armitage(2007)]{alex07} Alexander,~R.~D. \& Armitage,~P.~J. 2007, \mnras, 375, 500 
  \bibitem[Andrews \& Williams(2007)]{andr07} Andrews,~S.~M. \& Williams,~J.~P. 2007, \apj, 671, 1800 
  \bibitem[Andrews et~al.(2011)]{andr11} Andrews,~S.~M., Wilner,~D.~J., \& Espaillat,~C. et~al. 2011, \apj, 732, 42 
%  \bibitem[Andrews et~al.(2012)]{andr12} Andrews,~S.~M., Wilner,~D.~J., \& Hughes,~A.~M. et~al. 2012, \apj, 744, 162 
%  \bibitem[Artymowicz \& Lubow(1994)]{arty94} Artymowicz,~P. \& Lubow,~S.~H. 1994, \apj, 421, 651
  \bibitem[Ataiee et~al.(2013)]{atai13} Ataiee,~S., Pinilla,~P., \& Zsom,~A. et~al. 2013, \aap, 553, L3 
  \bibitem[Avenhaus et~al.(2014)]{aven14} Avenhaus,~H., Quanz,~S.~P., \& Meyer,~M.~R. et~al. 2014, \apj, 790, 56
  \bibitem[Balbus \& Hawley(1991)]{balb91} Balbus,~S.~A. \& Hawley,~J.~F. 1991, \apj, 376, 214 
  \bibitem[Baraffe et~al.(2003)]{bara03} Baraffe,~I. Chabrier,~G., Barman,~T.~S., Allard,~F., \& Hauschildt,~P.~H. 2003, \aap, 402, 701
  \bibitem[Barge \& Sommeria(1995)]{barg95} Barge,~P. \& Sommeria,~J. 1995, \aap, 295, L1 
  \bibitem[Beckwith et~al.(1990)]{beck90} Beckwith,~S.~V.~W., Sargent,~A.~I., Chini,~R.~S., \& Guesten,~R. 1990, \aj, 99, 924 
  \bibitem[Beckwith \& Sargent(1991)]{beck91} Beckwith,~S.~V.~W. \& Sargent,~A.~I. 1991, \apj, 381, 250 
%  \bibitem[Bessell, Castelli, \& Plez(1998)]{bess98} Bessell,~M.~S., Castelli,~F., \& Plez,~B. 1998, \aap, 333, 231
%  \bibitem[Beuzit et~al.(2008)]{beuz08} Beuzit,~J., Feldt,~M., \& Dohlen,~K. 2008, SPIE, 7014, 18
  \bibitem[Birnstiel, Andrews, \& Ercolano(2012)]{birn12} Birnstiel,~T., Andrews,~S.~M., \& Ercolano,~B. 2012, \aap, 544, 79
  \bibitem[Brown et~al.(2008)]{brow08} Brown,~J.~M., Blake,~G.~A., Qi,~C., Dullemond,~C.~P., \& Wilner,~D.~J. 2008, \apj, 675, L109 
  \bibitem[Brown et~al.(2009)]{brow09} Brown,~J.~M., Blake,~G.~A., \& Qi,~C. et al. 2009, \apj, 704, 496 
  \bibitem[Brown et~al.(2012)]{brow12} Brown,~J.~M., Herczeg,~G.~J., Pontoppidan,~K.~M., \& van~Dishoeck,~E.~F. 2012, \apj, 744, 116 
  \bibitem[Bruderer et~al.(2014)]{brud14} Bruderer,~S., van~der~Marel,~N., van~Dishoeck,~E.~F., \& van~Kempen,~T.~A. 2014, \aap, 562, A26 
%  \bibitem[Buenzli, Thalmann, \& Vigan(2010)]{buen10}  Buenzli,~E., Thalmann,~C., \& Vigan,~A. et al. 2010, \aap, 524, L1
  \bibitem[Calvet et~al.(2005)]{calv05} Calvet,~N., D'Alessio,~P., \& Watson,~D.~M. et~al. 2005, \apjl, 630, L185 
  \bibitem[Casassus et~al.(2013)]{casa13} Casassus,~S., van~der~Plas,~G.~M, \& Perez,~S. et~al. 2013, Nature, 493, 191
  \bibitem[Chiang et~al.(2001)]{chia01} Chiang,~E.~I., Joung,~M.~K., \& Creech-Eakman,~M.~J. et~al. 2001, \apj, 547, 1077 
  \bibitem[Clarke, Gendrin, \& Sotomayor(2001)]{clar01}  Clarke,~C., Gendrin,~A., \& Sotomayor,~M. 2001, \mnras, 328, 485
  \bibitem[Close et~al.(2014)]{clos14} Close,~L.~M., Follette,~K.~B., \& Males,~J.~R., et~al. 2014, \apj, 781, L30
%  \bibitem[Chiang \& Murray-Clay (2007)]{chia07} Chiang,~E. \& Murray-Clay,~R. 2007, Nat.~Phys., 3, 604
%  \bibitem[Chun et~al.(2008)]{chun08} Chun,~M. et al. 2008, Proc. SPIE, 7015, 49
%  \bibitem[Cieza et~al.(2012)]{ciez12} Cieza,~L.~A., Mathews,~G.~S., \& Williams,~J.~P. et~al. 2012, \apj, 752, 75 
  \bibitem[Cutri et~al.(2013)]{cutr13} Cutri,~R.~M., et al. 2013, Explanatory Supplement to the AllWISE Data Release Products, Tech. rep
  \bibitem[D'Alessio et~al.(1999)]{dale99} D'Alessio,~P., Calvet,~N., Hartmann,~L., Lizano,~S., \& Cant\'{o},~J. 1999, \apj, 527, 893
  \bibitem[D'Alessio et~al.(2001)]{dale01} D'Alessio,~P., Calvet,~N., \& Hartmann,~L. 2001, \apj, 553, 321 
  \bibitem[D'Alessio et~al.(2006)]{dale06} D'Alessio,~P., Calvet,~N., Hartmann,~L., Franco-Hern\'{a}ndez,~R., \& Serv\'{i}n,~H. 2006, ApJ, 638, 314 
  \bibitem[Debes et~al.(2013)]{debe13} Debes,~J.~H., Jang-Condell,~H., Weinberger,~A.~J., Roberge,~A., \& Schneider,~G. 2013, \apj, 771, 45 
  \bibitem[Dodson-Robinson \& Salyk (2011)]{dods11} Dodson-Robinson,~S. \& Salyk,~C. 2011, \apj, 738, 131 
  \bibitem[Dong et~al.(2012a)]{dong12a} Dong,~R., Rafikov,~R., \& Zhu,~Z. et~al. 2012, \apj, 750, 161 
  \bibitem[Dong et~al.(2012b)]{dong12b} Dong,~R., Hashimoto,~J., \& Rafikov,~R. et~al. 2012, \apj, 760, 111 
  \bibitem[Draine (2006)]{drai06} Draine,~B.~T. 2006, \apj, 636, 1114 
  \bibitem[de~Juan~Ovelar et~al.(2013)]{deju13} de~Juan~Ovelar,~M., Min,~M., Dominik,~C., et~al. 2013, \aap, 560, A111
%  \bibitem[Dullemond \& Dominik(2004a)]{dull04a} Dullemond, C. P. \& Dominik, C. 2004a, \aap, 417, 159
%  \bibitem[Dullemond \& Dominik(2004b)]{dull04b} Dullemond, C. P. \& Dominik, C. 2004b, \aap, 421, 1075
  \bibitem[Dullemond \& Dominik(2005)]{dull05} Dullemond, C. P. \& Dominik, C. 2005, \aap, 434, 971
%  \bibitem[Espaillat et~al.(2007)]{espa07} Espaillat,~C., Calvet,~N., \& D'Alessio,~P. et~al. 2007, \apj, 670, L135
%  \bibitem[Espaillat et~al.(2011)]{espa11} Espaillat,~C., Furlan,~E., \& D'Alessio,~P. et~al. 2011, \apj, 728, 49 
  \bibitem[Espaillat et~al.(2014)]{espa14} Espaillat,~C., Muzerolle,~J., Najita,~J., et~al. 2014, in Protostars and Planets VI, in press
%  \bibitem[Flaherty et~al.(2012)]{flah12} Flaherty,~K., Muzerolle,~J., \& Rieke,~G. et~al. 2012, \apj, 748, 71 
  \bibitem[Follette et~al.(2013)]{foll13} Follette,~K.~B., Tamura,~M., \& Hashimoto,~J. et~al. 2013, \apj, 767, 10 
  \bibitem[Fukagawa et~al.(2006)]{fuka06} Fukagawa,~M., Tamura,~M., Itoh,~Y., Kudo,~T., Imaeda,~Y., Oasa,~Y., Hayashi,~S.~S., \& Hayashi,~M. 2006, \apj, 636, L153
  \bibitem[Fukagawa et~al.(2013)]{fuka13} Fukagawa,~M., Tsukagoshi,~T., \& Momose,~M. et~al. 2013, \pasj, 65, L14 
  \bibitem[Furlan et~al.(2009)]{furl09} Furlan,~E., Watson,~D.~M., \& McClure,~M.~K. et~al. 2009, \apj, 703, 1964 
  \bibitem[Gammie (1996)]{gamm96} Gammie,~C.~F. 1996, \apj, 457, 355 
  \bibitem[Garufi et~al.(2013)]{garu13} Garufi,~A., Quanz,~S.~P., \& Avenhaus,~H. et~al. 2013, \aap, 560, A105 
  \bibitem[Geers et~al.(2007)]{geer07} Geers,~V.~C., Pontoppidan,~K.~M., \& van~Dishoeck,~E.~F. et~al. 2007, \aap, 469, L35 
%  \bibitem[Gregorio-Hetem et~al.(1992)]{greg92} Gregorio-Hetem,~J., Lepine,~J.~R.~D., Quast,~G.~R., Torres,~C.~A.~O., \& de~La~Reza,~R. 1992, \aj, 103, 549
  \bibitem[Grady et~al.(2013)]{grad13} Grady,~C.~A., Muto,~T., \& Hashimoto,~J. et~al. 2013, \apj, 762, 48 
  \bibitem[Gregorio-Hetem \& Hetem(2002)]{greg02} Gregorio-Hetem,~J., \& Hetem,~A. 2002, MNRAS, 336, 197
%  \bibitem[Goldreich \& Tremaine(1979)]{gold79} Goldreich,~P., \& Tremaine,~S. 1979, \apj, 233, 857 
  \bibitem[Hartmann et~al.(1998)]{hart98} Hartmann,~L., Calvet,~N., Gullbring,~E., \& D\'Alessio,~P. 1998, ApJ, 495, 385 
  \bibitem[Hashimoto et~al.(2011)]{hash11} Hashimoto,~J., Tamura,~M., \& Muto,~T. et al. 2011, \apjl, 729, L17 
  \bibitem[Hashimoto et~al.(2012)]{hash12} Hashimoto,~J., Dong,~R., \& Kudo,~T. et~al. 2012, \apjl, 758, L19 
%  \bibitem[Hayano et~al.(2004)]{haya04} Hayano,~Y. et al. 2004, \procspie, 5490, 1088
  \bibitem[Hayashi, Nakazawa, \& Nakagawa(1985)]{haya85} Hayashi,~C., Nakazawa,~K., \& Nakagawa,~Y. 1985, in Protostars and Planets II, ed. D.~C.~Black, \& M.~S.~Matthews (Tucson, AZ: Univ. Arizona Press), 1100
%  \bibitem[Ireland \& Kraus(2008)]{irel08} Ireland,~M.~J. \& Kraus,~A.~L. 2008, \apj, 678, L59
  \bibitem[Isella et al.(2013)]{isel13} Isella,~A., P\'{e}rez,~L.~M., \& Carpenter,~J.~M. et~al. 2013, \apj, 775, 30 
%  \bibitem[Jarrett et~al.(2011)]{jarr11} Jarrett,~T.~H. et~al. 2011, \apj, 735, 112
  \bibitem[Kim, Martin, \& Hendry(1994)]{kim94} Kim,~S.~-H., Martin,~P.~G., \& Hendry,~P.~D. 1994, ApJ, 422, 164 
  \bibitem[Kley \& Dirksen(2006)]{kley06} Kley,~W. \& Dirksen,~G. 2006, \aap, 447, 369 
  \bibitem[Kley \& Nelson(2012)]{kley12} Kley,~W. \& Nelson,~R.~P. 2012, \araa, 50, 211 
%  \bibitem[Kraus \& Ireland(2012)]{krau12} Kraus,~A.~L. \& Ireland,~M.~J. 2012, \apj, 745, 5
%  \bibitem[Lafreni\'{e}re et~al.(2007)]{lafr07} Lafreni\'{e}re,~D., Marois,~C., Doyon,~R., Nadeau,~D., \& Artigau,~E. 2007, \apj, 660, 770
  \bibitem[Li \& Draine(2001)]{li01} Li,~A. \& Draine,~B.~T. 2001, \apj, 554, 778 
%  \bibitem[Lin \& Papaloizou(1986)]{lin86} Lin,~D.~N.~C. \& Papaloizou,~J.~C.~B. 1986, \apj, 307, 395
  \bibitem[Li, Li, \& Koller (2005)]{li05} Li,~H., Li,~S., \& Koller,~J. et~al. 2005, \apj, 624, 1003 
  \bibitem[Lin \& Papaloizou(2010)]{lin10} Lin,~M.-K. \& Papaloizou,~J.~C.~B. 2010, \mnras, 405, 1473 
  \bibitem[Lovelace et~al.(1999)]{love99} Lovelace,~R.~V.~E., Li,~H., Colgate,~S.~A., \& Nelson,~A.~F. 1999, \apj, 513, 805 
  \bibitem[Lubow \& D'Angelo(2006)]{lubo06} Lubow,~S.~H. \& D'Angelo,~G. 2006, \apj, 641, L526 
  \bibitem[Lynden-Bell \& Pringle(1974)]{lynd74} Lynden-Bell,~D. \& Pringle,~J.~E. 1974, MNRAS, 168, 603 
  \bibitem[Lyo et al.(2011)]{lyo11} Lyo,~A.~-R., Ohashi,~N., Qi,~C., Wilner,~D.~J., \& Su,~Y.~-N. 2011, \aj, 142, 151 
%  \bibitem[Lin \& Papaloizou(1993)]{lin93} Lin,~D.~N.~C. \& Papaloizou,~J.~C.~B. 1993, in Protostars and Planets III, ed. E.~Levy \& M.~S.~Matthews (Tucson, AZ: Univ. Arizona Press), 749                       
  \bibitem[Marois et~al.(2006)]{maro06} Marois,~C., Lafreni\'{e}re,~D., Doyon,~R., Macintosh,~B., \& Nadeau,~D. 2006, \apj, 641, 556
%  \bibitem[Marois et~al.(2008)]{maro08} Marois,~C. et~al. 2008, Science, 322, 1348         
  \bibitem[Mathews, Williams, \& M\`{e}nard (2012)]{math12} Mathews,~G.~S., Williams,~J.~P., \& M\`{e}nard,~F. 2012, \apj, 753, 59 
%  \bibitem[Mathis (1990)]{math90} Mathis,~J.~S. 1990, ARA\&A, 28, 37
  \bibitem[Mayama et~al.(2012)]{maya12} Mayama,~S., Hashimoto,~J., \& Muto,~T. et~al. 2012, \apjl, 760, L26 
%  \bibitem[Metchev, Hillenbrand, \& Meyer(2004)]{metc04} Metchev, S. A., Hillenbrand, L. A., \& Meyer, M. R. 2004, \apj, 600, 435
%  \bibitem[Moshir (1989)]{mosh89} Moshir,~M. 1989, IRAS Faint Source Catalog, VizieR Online Data Catalog II/156A
  \bibitem[Muto et~al.(2012)]{muto12} Muto,~T., Grady,~C.~A., \& Hashimoto,~J. et~al. 2012, \apj, 748, L22 
%  \bibitem[Muzerolle et~al.(2009)]{muze09} Muzerolle,~J., Flaherty,~K., \& Balog,~Z. et~al. 2009, \apj, 704, L15 
  \bibitem[Najita, Strom, \& Muzerolle(2007)]{naji07} Najita,~J.~R., Strom,~S.~E., \& Muzerolle,~J. 2007, \mnras, 378, 369 
%  \bibitem[Nakajima et~al.(1989)]{naka89} Nakajima,~T. et~al. 1989, \aj, 97, 1510
  \bibitem[Okuzumi \& Hirose(2011)]{okuz11} Okuzumi,~S. \& Hirose,~S. 2011, \apj, 742, 65 
  \bibitem[Owen (2014)]{owen14} Owen,~J.~E. 2014, \apj, 789, 59O
  \bibitem[Papaloizou et~al.(2007)]{papa07} Papaloizou,~J.~C.~B., Nelson,~R.~P., Kley,~W., Masset,~F.~S., \& Artymowicz,~P. 2007, in Protostars and Planets V, ed. B.~Reipurth, D.~Jewitt, \& K.~Keil (Tucson: Univ. Arizona Press), 655
  \bibitem[P\'{e}rez et~al.(2014)]{pere14} P\'{e}rez,~L.~M., Isella,~A., Carpenter,~J.~M., \& Chandler,~C.~J. 2014, \apj, 783, L13
%  \bibitem[Perrin et~al.(2009)]{perr09} Perrin,~M.~D., Schneider,~G., Duchene,~G., Pinte,~C., Grady,~C.~A., Wisniewski,~J.~P., \& Hines,~D.~C. 2009, \apjl, 707, L132
  \bibitem[Pi\'{e}tu et~al.(2005)]{piet05} Pi\'{e}tu,~V., Guilloteau,~S., \& Dutrey,~A. 2005, \aap, 443, 945 
%  \bibitem[Pi\'{e}tu et~al.(2006)]{piet06} Pi\'{e}tu,~V., Dutrey,~A., Guilloteau,~S., Chapillon,~E., \& Pety,~J. 2006, \aap, 460, L43
  \bibitem[Pineda et~al.(2014)]{pine14} Pineda,~J.~E., Quanz,~S.~P., Meru,~F., Mulders,~G.~D., Meyer,~M.~R., Pani\'{c},~O., \& Avenhaus,~H. 2014, \apj, 788, 34
  \bibitem[Pinilla, Benisty, \& Birnstiel(2012)]{pini12} Pinilla,~P., Benisty,~M., \& Birnstiel,~T. 2012a, \aap, 545, A81 
  \bibitem[Quanz et~al.(2013)]{quan13} Quanz,~S.~P., Avenhaus,~H., \& Buenzli,~E. et~al. 2013, \apjl, 766, L2
  \bibitem[Racine et~al.(1999)]{raci99} Racine,~R., Walker,~G.~A.~H., Nadeau,~D., Doyon,~R., \& Marois,~C. 1999, \pasp, 111, 587 
  \bibitem[Reg\'{a}ly et~al.(2012)]{rega12} Reg\'{a}ly,~Z., Juh\'{a}sz,~A., S\'{a}ndor,~Z., \& Dullemond,~C.~P. 2012, \mnras, 419, 1701 
  \bibitem[Reg\'{a}ly, Kir\'{a}ly, \& Kiss(2014)]{rega14} Reg\'{a}ly,~Zs., Kir\'{a}ly,~S., \& Kiss,~L.~L. 2014, \apj, 785, 31
  \bibitem[Riaud et~al.(2006)]{riau06} Riaud,~P., Boccaletti,~A., Baudrand,~J., \& Rouan,~D. 2006, \aap, 458, 317
  \bibitem[Rice et~al.(2006)]{rice06} Rice,~W.~K.~M., Armitage,~P.~J., Wood,~K., \& Lodato,~G. 2006, \mnras, 373, 1619 
%  \bibitem[Shu, Adams, \& Lizano(1987)]{shu87} Shu,~F.~H., Adams,~F.~C., \& Lizano,~S. 1987, ARA\&A, 25, 23      
%  \bibitem[Roeser, Demleitner, \& Schilbach(2010)]{rose10} Roeser,~S., Demleitner,~M., \& Schilbach,~E. 2010, \aj, 139, 2440
  \bibitem[Sano et~al.(2000)]{sano00} Sano,~T., Miyama,~S.~M., Umebayashi,~T., \& Nakano,~T. 2000, \apj, 543, 486 
  \bibitem[Sault, Teuben, \& Wright(1995)]{saul95} Sault,~R.~J., Teuben,~P.~J., \& Wright,~M.~C.~H. 1995, ASP Conf. Ser. 77, Astronomical Data Analysis Software and Systems IV, ed. R.~A.~Shaw, H.~E.~Payne, \& J.~J.~E.~Hayes (San Francisco, CA: ASP), 433 
%  \bibitem[Suzuki, Muto, \& Inutsuka(2010)]{suzu10} Suzuki,~T.~K., Muto,~T., \& Inutsuka,~S.-i. 2010, \apj, 718, 1289
  \bibitem[Strom et~al.(1989)]{stro89} Strom,~K.~M., Strom,~S.~E., Edwards,~S., Cabrit,~S., \& Skrutskie,~M.~F. 1989, \aj, 97, 1451
%  \bibitem[Tamura et~al.(2006)]{tamu06} Tamura,~M. et~al. 2006, \procspie, 6269, 28
%  \bibitem[Tamura (2009)]{tamu09} Tamura,~M. 2009, in proc. ``Exoplanetsand Disks: Their Formation and Diversity'' (2009 March9-12), eds. Usuda, T., Ishii, M. and Tamura, M. (National Astronomical Observatory of Japan), p. 11-16
  \bibitem[Takami et~al.(2014)]{taka14} Takami,~M., Hasegawa,~Y., \& Muto,~T. et~al. 2014, \apj, ApJ, 795, 71
  \bibitem[Tang et~al.(2012)]{tang12} Tang,~Y.~-W., Guilloteau,~S., \& Pi\'{e}tu,~V. et~al. 2012, \aap, 547, A84 
  \bibitem[Terquem (2008)]{terq08} Terquem,~C.~E.~J.~M.~L.~J. 2008, \apj, 689, 532
  \bibitem[Thalmann et~al.(2010)]{thal10} Thalmann,~C., Grady,~C.~A., \& Goto,~M. et~al. 2010, \apj, 718, L87 
  \bibitem[Tsukagoshi et~al.(2014)]{tsuk14} Tsukagoshi,~T., Momose,~M., \& Hashimoto,~J. et~al. 2014, \apj, 783, 90
  \bibitem[van~der~Marel et~al.(2013)]{vand13} van~der~Marel,~N., van~Dishoeck,~E.~F., \& Bruderer,~S. et al. 2013, Sci, 340, 1199 
  \bibitem[Varni\'{e}re \& Tagger(2006)]{varn06} Varni\'{e}re,~P. \& Tagger,~M. 2006, \aa, 446, L13
  \bibitem[Verhoeff et~al.(2011)]{verh11} Verhoeff,~A.~P., Min,~M., \& Pantin,~E. et al. 2011, \aap, 528, A91
 % \bibitem[Tokunaga \& Vacca(2005)]{toku05} Tokunaga,~A.~T., \& Vacca,~W.~D. 2005, PASP, 117, 1459      
%  \bibitem[Weinberger et~al.(1999)]{wein99} Weinberger,~A.~J., Becklin,~E.~E., Schneider,~G., Smith,~B.~A., Lowrance,~P.~J., Silverstone,~M.~D., Zuckerman,~B., \& Terrile,~R.~J. 1999, \apj, 525, L53                                 
  \bibitem[Williams \& Cieza(2011)]{will11} Williams,~J.~P. \& Cieza,~L.~A. 2011, ARA\&A, 49, 67
  \bibitem[Whitney et~al.(2013)]{whit13} Whitney,~B.~A., Robitaille,~T.~P., \& Bjorkman,~J.~E. et~al. 2013, ApJS, 207, 30 
  \bibitem[Wood et~al.(2002)]{wood02} Wood,~K., Wolff,~M.~J., Bjorkman,~J.~E., \& Whitney,~B. 2002, \apj, 564, 887 
  \bibitem[Zhu et~al.(2011)]{zhu11} Zhu,~Z., Nelson,~R.~P., Hartmann,~L., Espaillat,~C., \& Calvet,~N. 2011, \apj, 729, 47
  \bibitem[Zhu et~al.(2012)]{zhu12} Zhu,~Z., Nelson,~R.~P., Dong,~R., Espaillat,~C., \& Hartmann,~L. 2012, \apj, 755, 6 
  \bibitem[Zhu et~al.(2014)]{zhu14a} Zhu,~Z., Stone,~J.~M., Rafikov,~R.~R. \& Bai,~X. 2014, \apj, 785, 122
  \bibitem[Zhu \& Stone(2014)]{zhu14b} Zhu,~Z. \& Stone,~J.~M. 2014, \apj, \apj, 795, 53
    
  \end{thebibliography}
\end{document}